\documentclass[twocolumn]{aastex701}
\usepackage{graphics,graphicx}
\hypersetup{urlcolor=blue}
\usepackage{natbib}
\usepackage[outdir=./]{epstopdf}
\usepackage{textcomp,gensymb}
\usepackage{xfrac}
\usepackage{xcolor}
\usepackage{multirow}
\usepackage{siunitx}

\newcommand{\kepler}{{\it Kepler}}

\newcommand{\gaia}{{\textit {Gaia}}}

\usepackage[outdir=./]{epstopdf}
\usepackage{ulem}
\usepackage{mathrsfs}
\usepackage{fontawesome}
\usepackage{comment}

\newcommand{\tess}{\textit{TESS}}

\newcommand{\spitzer}{\textit{Spitzer}}

\newcommand{\jwst}{\textit{JWST}}
\newcommand{\plnameb}{HIP\,67522\,b}
\newcommand{\starname}{HIP\,67522}
\newcommand{\plnamec}{HIP\,67522\,c}

\newcommand{\cheops}{\textit{CHEOPS}}
\newcommand{\btransit}{55}
\newcommand{\ctransit}{29}
\newcommand{\totaltransits}{84}
\newcommand{\bepochs}{50}
\newcommand{\cepochs}{22}

\newcommand{\unc}{Department of Physics and Astronomy, The University of North Carolina at Chapel Hill, Chapel Hill, NC 27599, USA}
\newcommand{\abc}{Astrobiology Center, 2-21-1 Osawa, Mitaka, Tokyo 181-8588, Japan}
\newcommand{\naoj}{National Astronomical Observatory of Japan, 2-21-1 Osawa, Mitaka, Tokyo 181-8588, Japan}
\newcommand{\sokendai}{Department of Astronomy, School of Science, The Graduate University for Advanced Studies (SOKENDAI), 2-21-1 Osawa, Mitaka, Tokyo, Japan}

\newcommand{\iac}{Instituto de Astrof\'\i sica de Canarias (IAC), 38205 La Laguna, Tenerife, Spain}

\newcommand{\komabasc}{Komaba Institute for Science, The University of Tokyo, 3-8-1 Komaba, Meguro, Tokyo 153-8902, Japan}

\pdfoutput=1

\shorttitle{TTVs in HIP\,67522} 
\shortauthors{James et al. 2026}
\submitjournal{The Astronomical Journal}

\begin{document}

\title{Toddlers in Resonance I:\\ low masses and eccentricities of the 17\,Myr giant planets HIP\,67522\,bc from transit timing variations} 

\author[0009-0005-8083-3954]{Hannah R. James}
\affiliation{\unc} 
\email{hanjam@unc.edu}

\author[0000-0002-8399-472X]{Madyson G. Barber}
\altaffiliation{NSF Graduate Research Fellow}
\affiliation{\unc} 
\email{madysonb@live.unc.edu }

\author[0000-0003-3654-1602]{Andrew W. Mann} 
\affiliation{\unc}
\email{awmann@unc.edu}

\author[0000-0001-5729-6576]{Pa Chia Thao} 
\affiliation{\unc} 
\email{pachia@live.unc.edu}

\author[0000-0001-9811-568X]{Adam L. Kraus} 
\affiliation{Department of Astronomy, The University of Texas at Austin, Austin, TX 78712, USA}
\email{alk@astro.as.utexas.edu}

\author[0000-0003-2053-07492]{Benjamin M. Tofflemire} 
\altaffiliation{51 Pegasi b Fellow}
\affiliation{Department of Astronomy, The University of Texas at Austin, Austin, TX 78712, USA}
\email{btofflemire@seti.org}

\author[0000-0001-7246-5438]{Andrew Vanderburg} 
\affiliation{Center for Astrophysics \textbar \ Harvard \& Smithsonian, 60 Garden Street, Cambridge, MA 02138, USA}
\email{andrew.m.vanderburg@gmail.com}

\author[0000-0002-0856-4527]{Lyu Abe} 
\affiliation{Université Côte d'Azur, Observatoire de la Côte d'Azur, CNRS, Laboratoire Lagrange, CS 34229, F-06304 Nice Cedex 4, France}
\email{Lyu.Abe@oca.eu}

\author[0000-0002-7188-8428]{Tristan Guillot} 
\affiliation{Université Côte d'Azur, Observatoire de la Côte d'Azur, CNRS, Laboratoire Lagrange, CS 34229, F-06304 Nice Cedex 4, France}
\email{tristan.guillot@oca.eu}

\author[0000-0002-3503-3617]{Olga Suarez} 
\affiliation{Université Côte d'Azur, Observatoire de la Côte d'Azur, CNRS, Laboratoire Lagrange, CS 34229, F-06304 Nice Cedex 4, France}
\email{olga.suarez@oca.eu}

\author[0000-0001-5000-7292]{Djamel Mekarnia} 
\affiliation{Université Côte d'Azur, Observatoire de la Côte d'Azur, CNRS, Laboratoire Lagrange, CS 34229, F-06304 Nice Cedex 4, France}
\email{djamel.mekarnia@oca.eu}

\author[0000-0002-5510-8751]{Amaury H. M. J. Triaud} 
\affiliation{School of Physics \& Astronomy, University of Birmingham, Edgbaston, Birmingham B15 2TT, UK}
\email{a.triaud@bham.ac.uk}

\author[0009-0007-5876-546X]{Vincent Deloupy} 
\affiliation{École Normale Supérieure, Département de Physique, Rue d'Ulm, 75005 Paris Cedex 5, France}
\email{vincent.deloupy@gmail.com}

\author[0000-0003-1368-6593]{Mayuko Mori} 
\affiliation{\abc}
\affiliation{\naoj}
\email{mayukomori.519@gmail.com}

\author[0000-0002-6424-3410]{Jerome P. de Leon} 
\affiliation{\komabasc}
\email{jpdeleon@g.ecc.u-tokyo.ac.jp}

\author[0000-0001-8511-2981]{Norio Narita} 
\affiliation{\komabasc}
\affiliation{\iac}
\affiliation{\abc}
\email{narita@g.ecc.u-tokyo.ac.jp}

\author[0000-0002-4909-5763]{Akihiko Fukui} 
\affiliation{\komabasc}
\affiliation{\iac}
\email{afukui@g.ecc.u-tokyo.ac.jp}

\author[0000-0002-4881-3620]{John H. Livingston} 
\affiliation{\abc}
\affiliation{\naoj}
\affiliation{\sokendai}
\email{john.livingston@nao.ac.jp}

\author[0000-0002-4891-3517]{George Zhou} 
\affiliation{University of Southern Queensland, Centre for Astrophysics, West Street, Toowoomba, QLD 4350 Australia}
\email{George.Zhou@unisq.edu.au}

\author[0009-0001-7756-9401]{Camille M. L. Guenée} 
\affiliation{Université Claude Bernard Lyon 1, Département de Physique, UFR Faculté des Sciences, F-69622 Villeurbanne, France} \email{camille.guenee@gmail.com}

\author[0000-0002-7733-4522]{Juliette Becker} 
\affiliation{Department of Astronomy, University of Wisconsin–Madison, 475 N Charter St, Madison, WI 53706, USA}
\email{juliette.becker@wisc.edu}

\author[0009-0006-9572-1733]{Ana Isabel Lopez Murillo} 
\affiliation{\unc} 
\email{isaana@email.unc.edu}

\author[0000-0001-6588-9574]{Karen A.\ Collins} 
\affiliation{Center for Astrophysics \textbar \ Harvard \& Smithsonian, 60 Garden Street, Cambridge, MA 02138, USA}
\email{karen.collins@cfa.harvard.edu}

\author[0000-0001-8227-1020]{Richard P. Schwarz} 
\affiliation{Center for Astrophysics \textbar \ Harvard \& Smithsonian, 60 Garden Street, Cambridge, MA 02138, USA}
\email{rpschwarz@comcast.net}

\author[0000-0003-2163-1437]{Chris Stockdale} 
\affiliation{Hazelwood Observatory, Australia}
\email{thestockdalefamily@bigpond.com}

\author[0000-0001-5603-6895]{Thiam-Guan Tan} 
\affiliation{Perth Exoplanet Survey Telescope, Perth, Western Australia}
\email{tgtan@bigpond.net.au}

\author[0000-0003-1728-0304]{Keith Horne} 
\affiliation{SUPA Physics and Astronomy, University of St. Andrews, Fife, KY16 9SS Scotland, UK}
\email{kdh1@st-andrews.ac.uk}

\author[0009-0008-0940-1317]{Leah J. Boff} 
\email{lboff@unc.edu}
\altaffiliation{UNC Chancellor's Science Scholar}
\affiliation{\unc} 
\email{lboff@unc.edu}

\author[0000-0001-7336-7725]{Mackenna L. Wood} 
\affiliation{Museum of Science \& Innovation, 4801 E Fowler Ave, Tampa, FL 33617}
\email{mackenna.wood@MOSI.ORG}

\author[0000-0003-2535-3091]{Nidia Morrell} 
\affiliation{Las Campanas Observatory, Carnegie Observatories, Casilla 601, La Serena, Chile}
\email{nmorrell@carnegiescience.edu}

\author[orcid=0000-0002-0514-5538]{Luke G. Bouma} 
\affiliation{Observatories of the Carnegie Institution for Science, Pasadena, CA 91101, USA}
\affiliation{Caltech/IPAC, Pasadena, CA 91125, USA}
\email{lbouma@ipac.caltech.edu}

\begin{abstract} 
Infant ($<$100\,Myr) transiting planet systems provide glimpses into the processes that shape the observed mature planet population. Masses for these planets are powerful data points for testing theories of planet formation and evolution. Because young systems tend to be in mean-motion resonances (MMR) and show higher rates of transit timing variations (TTVs), we can take advantage of deviations in the expected versus observed transit timings due to gravitational interactions between the planets to determine the planetary masses and orbital eccentricities. Here we report the TTV analysis of HIP\,67522, a 17\,Myr two-planet system near a 2:1 orbital resonance.  
The analysis is based on \btransit\ transit observations (\bepochs\ epochs) of planet b and \ctransit\ transit observations (\cepochs\ epochs) of planet c, taken across 11 facilities. Assuming circular orbits, we determine masses of $M_b=25.0^{+7.6}_{-7.8}M_\oplus$ and $M_c=11.2\pm1.4M_\oplus$. Allowing for eccentric orbits, we find $M_b=23.1^{+15.4}_{-12.4}M_\oplus$ and $M_c=11.8^{+3.0}_{-2.3}M_\oplus$, and low eccentricities ($<0.1$) for both planets. Adopting a Gaussian prior on $M_b$ from the JWST transmission spectrum tightens the eccentric-solution mass for planet c to $9.7^{+1.9}_{-1.6}M_\oplus$, which is our recommended solution. The low masses and large radii are consistent with a contracting atmosphere planet formation model, such as Gas Dwarf formation. Given the challenge of radial-velocity masses and the high rate of TTVs seen in young systems, TTV-masses will be essential for testing theories of the formation and evolution of super-Earths and sub-Neptunes and determining the conditions that create the most common types of planets in the Galaxy.
\end{abstract}

\keywords{}

\section{Introduction}\label{sec:intro}

The discovery and characterization of young planetary systems ($\lesssim 100$ Myr) provide a direct window into the initial conditions of planetary architectures and the mechanisms that drive their evolution. These systems allow us to distinguish between various evolutionary pathways such as photoevaporation, core-powered mass loss, and orbital migration \citep[e.g.,][]{OwenWu2013,Mann2016b, GinzburgCorepowered2018}, test theories of planet formation \citep{Bean2021, Rogers2025}, and probe the initial atmospheres of the most common types of planets \citep{Thao2023, Barat2025_jwst}.

Measuring the masses of these young planets is critical for calculating their bulk densities and constraining their atmospheric compositions. Mass measurements for young planets can also constrain their initial entropy \citep{Owen2020}, and because young planets have inflated radii \citep{Mann2017a,Vach2024}, masses are required to compare properties of planets across ages. Unfortunately, high levels of stellar activity intrinsic to young stars often overwhelms the radial velocity (RV) signals of the planet. The result is that numerous mass measurements in the literature \citep[e.g.,][]{SuarezMascareno2021, Zakhozhay2022} were later refuted \citep[e.g.][]{Blunt2023, del2026longest}. Many young planets detected only through RVs remain controversial \citep{Donati2020, Damasso2020}. Carefully designed large RV programs of young transiting systems have yielded masses \citep{Barragan2019, Barragan2022, Mallorquin2023}, although many have only yielded upper limits on their masses \citep{Benatti2021, Donati2025}. 

Recent surveys have revealed that young planets are preferentially found in or near mean-motion resonances (MMR) \citep{Dai2024}, a configuration likely resulting from smooth Type-I migration within the protoplanetary disk \citep{Pichierri2019, Izidoro2017}. These resonant chains usually break up through post-disk dynamical instabilities \citep{Izidoro2021, Choksi2026}, but while intact, provide a high occurrence rate of detectable Transit Timing Variations \citep[TTVs;][]{Agol2005, Holman2005}. Indeed, young planets exhibit a much higher rate of TTVs than their older counterparts \citep{LopezMurillo2026}. Given the challenges of RV measurements, these TTVs provide a promising route to measure the masses of a statistical sample of young planetary systems. This spawned the Toddlers in Resonance program. 

HIP\,67522\,b (TIC\,166527623; HD\, 120411; TOI-6551), a low-density Jovian-sized planet orbiting a $\sim 17$\,Myr-old solar analog in the Scorpius-Centaurus association \citep{THYMEII}, has emerged as a key test case for young planet characterization. \citet{Thao2024_featherweight} recently utilized transmission spectroscopy to infer a mass of $13.8\pm1.0$\,$M_{\oplus}$ for planet b, based on the amplitude of atmospheric features \citep{deWit2013}. Notably, \citet{Thao2024_featherweight} also reported a significant TTV signal in the TESS and CHEOPS photometry of planet b, which led to the confirmation of HIP\,67522\,c, a second transiting planet in the system \citep{Barber2024a} just outside a 2:1 period resonance.

However, the mass of HIP\,67522\,b remains a subject of active discussion. A recent analysis by \citet{Lavail2026} used transmission spectra from VLT/CRIRES+ to measure a significantly higher mass ($27.7^{+5.9}_{-5.5}\,M_\oplus$). This tension highlights a broader concern within the community regarding masses derived from transmission spectra: atmospheric properties such as high-altitude aerosols (clouds/hazes), complex abundances, non-isothermal chemistry, or errors in the input opacities can alter spectral features, leading to a potential over- or under-estimation of the planetary scale height and, consequently, its gravity \citep{Batalha2017, Welbanks2019, Ohno2020}. 

The retrieval run in \citet{Thao2024_featherweight} accounts for some of these complications, but uncertainties from input opacities (for example) are challenging to marginalize over. Scale-height masses, which would be invaluable for characterizing young puffy planets, would be considered more reliable if independently verified with TTVs. This is the case for V1298 Tau\,b, which has a mass from its transmission spectrum \citep{Barat2025_jwst} and a consistent mass from the system's TTVs \citep{Livingston2026}. In addition to providing a second benchmark system, TTVs for HIP\,67522 would provide a mass for the outer planet and constraints on the eccentricity (dynamical state) of both planets.

In this work, we combine data from eleven different facilities to perform a comprehensive TTV analysis of the HIP\,67522 system to provide a dynamical measurement of the masses and eccentricities of both planets b and c. In Section\,\ref{sec:obs}, we describe the new transit photometry and detail our data reduction procedure. Section\,\ref{sec:transitfitting} presents our transit fitting methodology and the derivations of individual mid-transits. In Section\,\ref{sec:ttv}, we detail our TTV analysis, the results of which we present in Section\,\ref{sec:results}. Finally, in Section\,\ref{sec:summary}, we summarize our findings and discuss their implications for the dynamical architecture of the system, as well as highlighting how planet c is an  excellent target for future transmission spectroscopy with \jwst.

\section{Observations and Data Reduction}\label{sec:obs}

We collected transit observations of \starname{} from eleven different facilities; \tess, \cheops, \spitzer, \jwst, SOAR, LCOGT, ASTEP, Minerva-Australis, Swope, NGTS, and PEST. Some of these transits were analyzed and published in prior studies of HIP\,67522 \citep[e.g.,][]{THYMEII,Thao2024_featherweight,Chakraborty2026}, but many are new datasets published here for the first time (e.g., data from ASTEP, SOAR, most of the LCOGT data, and the most recent \tess\ sectors) or the transits were not specifically analyzed in the prior papers (e.g., the \cheops\ data). With the only exception being the NGTS transit data (see Section\,\ref{subsec:ngts_data}), we reanalyzed the original transit data, as the more precise transit parameters helped provide more accurate and precise transit times. In total, we have \btransit{} transits for planet b and \ctransit{} transits for planet c, spanning more than seven years (2019-2026). Table\,\ref{tab:obslog} summarizes all transit observations used in this work.

\subsection{\tess}\label{sec:tess}
\starname\  was first observed by the Transiting Exoplanet Survey Satellite \citep[\tess\,;][]{Ricker2014} in Sector 11, which was observed from 2019\,Apr\,22 to May\,21. The target was pre-selected for short cadence because of its membership in Sco-Cen (G011280; PI Rizzuto). Subsequently, \starname\,  was re-observed in Sector 38, 64, 101, and 102. In total, there were 20 transits of planet b, and 8 transits of planet c in the \tess{} data.

The data were processed by the Science Processing and Operations Center (SPOC) pipeline \citep{jenkins2016tess}. The SPOC default processing sometimes struggles with rapid rotators due to the stellar signal occurring on similar timescales to the instrumental systematics, so we employed a custom extraction pipeline as outlined in \citet{Vanderburg2019} and \cite{Barber2024_iras}. This approach includes a simultaneous fitting of stellar variability with the co-trending basis vectors to improve the instrumental systematics model. This approach has worked well on prior young planetary systems with \tess\ \citep[e.g.,][]{THYMEIII, Thao2024_1224}. 

We estimated the flux uncertainties for each sector separately, adopting the median of three estimations: 1) the median absolute deviation of the point minus the adjacent point, 2) the median absolute deviation of the flattened light curve (flattened using \texttt{lightkurve} \citep{lightkurve} with a 4$\sigma$ outlier threshold), and 3) the standard deviation of the flattened light curve.

\subsection{CHEOPS}\label{subsec:cheops}

We drew additional transit observations of the HIP\,67522 system taken with the CHaracterising ExOPlanets Satellite \citep[\cheops;][]{Benz2021}. The observations span approximately 15 months (2024\,Mar\,09 to 2025\,Jun\,25) and were acquired from two \cheops\ Guest Observer programs. These \cheops\ data were previously published by \citet{Ilin2025} as part of an effort to characterize planet-induced stellar flares in the system. 

The dataset included four full transits of HIP\,67522\,b from program AO-4 ID17 (PI: H. Chakraborty) and a larger set of partial transits under program AO-4 ID4 (PI: E. Ilin). From this second program, we selected the 11 partial transits that captured sufficient fraction of the transit chord (e.g., including at least one distinct ingress or egress) to enable robust geometric constraints and timing measurements. In total, we analyzed 15 transit events. Notably, one of the partial transit visits for planet b overlaps with a partial transit of the outer planet, HIP\,67522\,c, which we model simultaneously (see Section\,\ref{sec:transitfitting}). 

The raw \cheops\ images were processed by the official \cheops\ Data Reduction Pipeline \citep[DRP v14.1.3;][]{Hoyer2020}, which automatically performs environmental monitoring, instrumental corrections (including bias, dark, and flat-fielding), and background sky estimation. We extracted the 1D photometric time series using the \texttt{pycheops} software package \citep{Maxted2023}, utilizing the default aperture radius of 25 pixels and applying the internal \texttt{pycheops} field-star decontamination algorithm to account for background flux dilution.

\subsection{Spitzer}

The \textit{Spitzer} Space Telescope took time-series observations of \starname\,b on 2019\,Dec\,09 as part of a program to validate young planetary systems (program ID 14011, PI Newton) with the Infrared Array Camera \citep[4.5{\textmu}m channel;][]{Fazio2004}. Data were reduced using the Photometry for Orbits, Eccentricities, and Transits pipeline \citep[POET;][]{Stevenson2012}. Significantly more details on the construction of this light curve can be found in the discovery paper \citet{THYMEII}.

\subsection{JWST}

\citet{Thao2024_featherweight} obtained a single transit of \starname\,b on 2023\,Feb\,26 UTC using the Near Infrared Spectrograph (NIRSpec) in the Bright Object Time Series (BOTS) mode on \jwst. This observation was part of a Cycle 1 program to characterize the atmosphere (GO: 2498; PI: A. Mann). As we discuss further in Section\,\ref{sec:summary}, the {\it JWST} data missed egress due to the uncertainty in early estimates of the period (in part due to the TTV). However, the precision of the data still provided extremely precise constraints on the timing. Details of the observations and reduction are given in \citet{Thao2024_featherweight}. For this paper, we used the two `white-light' curves from the NIRSpec detectors (NRS1 and NRS2) rather than the full spectroscopic curve. 

\subsection{SOAR} \label{sec:SOAR}

We observed three transits of HIP\,67522\,b using the Goodman High Throughput Spectrograph \citep{Goodman_spectrograph} attached to the 4.1-m Southern Astrophysical Research Telescope (SOAR) atop Cerro Pachon, Chile, in imaging mode. Conditions were near-photometric (thin or no clouds) on all three nights. 

All observations used the red camera with readout set to 750 Hz ATTN 0. In this mode, Goodman has a default \ang{;7.2;}-diameter circular field of view with a pixel scale of 0.\SI{15}{\arcsecond} pixel$^{-1}$. We used a small defocus (roughly 2-3 times the focused full-width half-max) to avoid saturation. To improve the readout speed, we tried using a subarray with a height of 1500 pixels for 2022\,May\,03, but found that the benefit was small compared to loss of comparison stars; we used a full or near-full readout for other nights. 

Basic reduction of the Goodman data followed that of \citet{THYMEVI} and \cite{Thao2024_featherweight}. We used the optimal aperture radius, which varied from 40 pixels to 110 pixels between observations, and a 40-pixel width sky annulus separated from the aperture boundary by at least 20 pixels. We built a master comparison light curve using a weighted mean of the 20 brightest stars consistently detected across all images (with apertures not touching the edge). To calculate the weights, we employed a leave-one-out (LOO) validation method. For each candidate comparison star $C_i$, we constructed a temporary master calibration curve using the weighted average of all other available comparison stars in the field. This master curve was then used to correct $C_i$. We calculated the precision of the resulting corrected light curve by measuring the Median Absolute Deviation (MAD) of the successive differences; the weights were the inverse-square of these precision estimates. 

\begin{figure}
    \centering
    \includegraphics[width=0.48\textwidth]{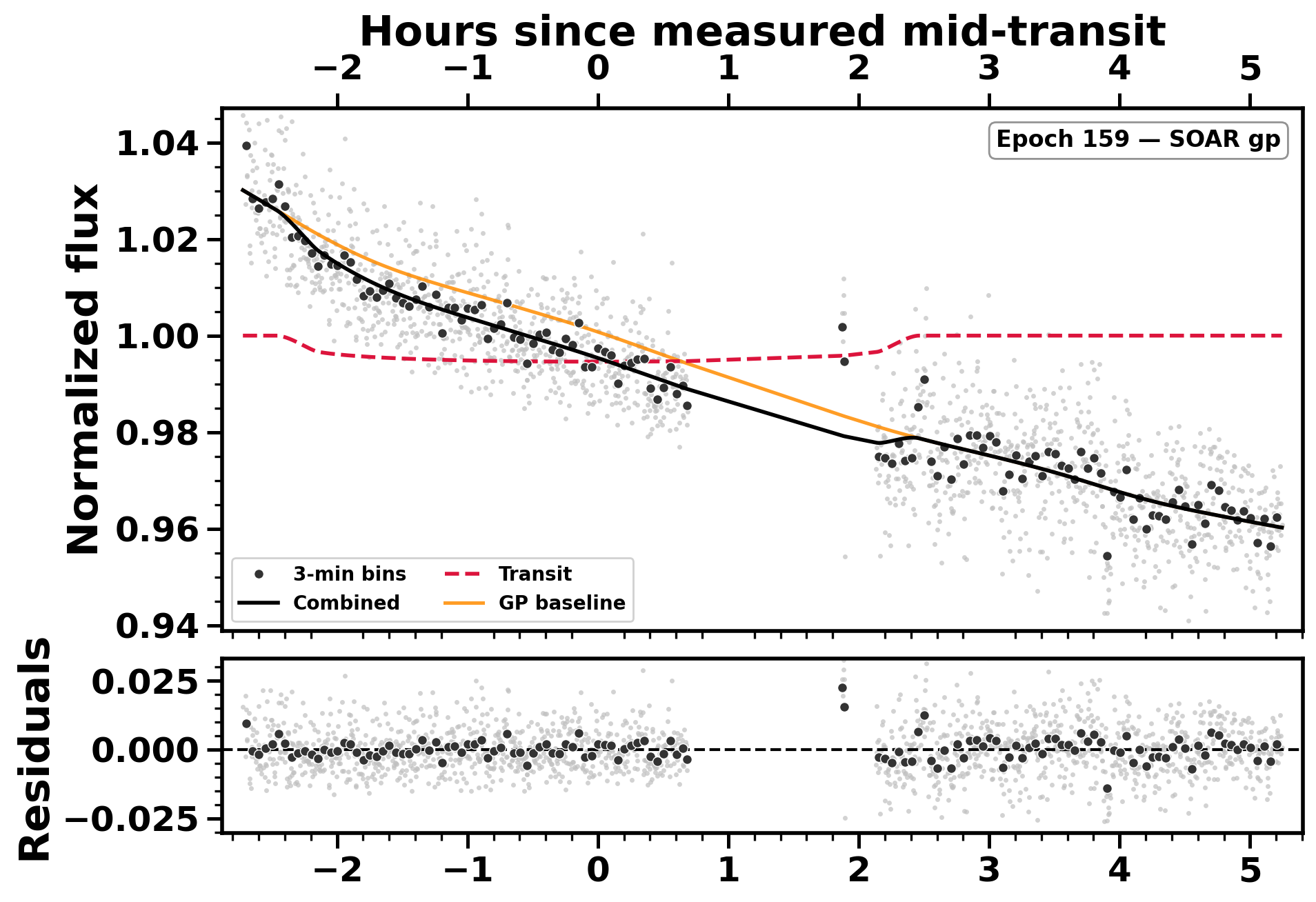}
    \caption{Excluded SOAR transit of HIP\,67522\,b. For clarity, we show 3-min bins of the photometry. There is a noticeable gap in the center of the data due to the elevation limits of SOAR in 2022 (post-earthquake). Acquisition was just before ingress and re-acquisition just before egress, meaning the transit timing was sensitive to how we extracted and calibrated the data. We opted to remove this transit from our analysis. }
    \label{fig:soar} 
\end{figure}

During the two 2022 transits, SOAR experienced guiding restrictions near zenith due to damage from the earthquake earlier that year. As a result, both transits had gaps lasting $\simeq$1\,hr while the target was at excluded elevations. Because of this, the second transit (see Figure\,\ref{fig:soar}) proved to be unreliable. The initial acquisition was close to ingress, and the post-guiding-loss re-acquisition was close to egress. The result is that there was degeneracy between the extraction and the transit time was extreme compared to all other data (15-30\,min off). Indeed, our initial fit of this dataset yielded a transit time that was $\simeq20$\,m early, a major reason \citet{Thao2024_featherweight} overestimated the amplitude of the TTV. We opted to remove this transit from our analysis. We kept the other two transits, but note that these yielded worse times than comparable ground-based data.

\subsection{LCOGT}

In total, we used 18 datasets across 12 epochs (5 of planet b, 7 of planet c) taken with telescopes from the Las Cumbres Observatory (LCOGT) network \citep{Brown13}. Several additional transits of each system were excluded due to poor coverage, weather, or large flares, which made the data unusable for timings. These were taken with a mix of LCOGT 0.4m-class telescopes and the SBIG or QHY camera, the 1m-class telescopes with the Sinistro camera \citep{Brown2011}, or the 2m-class telescopes with MuSCAT4 \citep{Narita2020}. The filter and exposure time varied, as not all of these were taken for the same science case (e.g., validation of the two planets, checking for spot crossings, and searching for TTVs). 

For each observation, we started with the standard LCOGT \texttt{BANZAI} Pipeline \citep{McCully18} calibrated images. For all of the 0.4m and most of the 1.0m photometry, we performed aperture photometry using \texttt{Photutils}, adopting circular apertures with radii ranging from 5--14 pixels, dependent on telescope, seeing conditions, and filter. We tested combinations of 3--10 comparison stars in the field with similar brightnesses, adopting the extraction that minimizes the point-to-point scatter. Similar to the \tess\ extraction, we estimate the uncertainty on the flux by taking the median of three estimations: 1) the median absolute deviation of the point minus the adjacent point, 2) the median absolute deviation of the flattened light curve (flattened using \texttt{lightkurve} \citep{lightkurve} with a 4$\sigma$ outlier threshold), and 3) the standard deviation of the flattened light curve.

For the three transits observed with the 2\,m MuSCAT4 imager \citep{Narita2020} listed in Table~\ref{tab:obslog}, as well as four of the 1.0m Sinistro observations, we performed a separate analysis. As before, we started from \texttt{BANZAI}-calibrated images. Photometry was performed with an independent, \texttt{prose}-based \citep{Garcia2022} pipeline developed for this work: for each band, a quality-vetted reference frame was used to detect sources and identify the target via WCS cross-match, aperture radii were scaled to the reference-frame FWHM and capped by a sky annulus sized to exclude Gaia neighbors contributing $\geq10\%$ of the target flux ($\Delta G < 2.5$~mag), and photometry was extracted in parallel across the full time series. Rather than a manual grid search over comparison-star combinations, differential photometry used the automated ensemble algorithm of \citet{Broeg2005}, which iteratively weights all non-flagged comparison stars in the field. Per-point flux uncertainties were propagated analytically from a CCD noise model (source Poisson noise, sky background, and read-noise/gain terms within the photometric aperture), rather than the empirical scatter-based estimator. 

As a check, we analyzed several 1.0m transits with both approaches (\texttt{Photutils} and \texttt{prose}) and found negligible changes in the resulting light curve. \texttt{prose} yielded slightly lower point-to-point scatter, but yielded trivial changes in the timings. Because these were so similar, we did not homogenize the LCOGT light curve extraction process.

During one night of LCOGT observations (2022\,May\,04) the target intermittently saturated the detector. However, the transit was still robustly detected using the non-saturated wings of the PSF or simply ignoring the saturated images. Nonlinearities still left significant structure (wiggles) in the extracted light curve, but ingress and egress were clear so the timing was still reliable. We opted to keep this transit.

\subsection{ASTEP}

The Antarctica Search for Transiting ExoPlanets (ASTEP) \citep{Guillot2015} is a 0.4\,m telescope equipped with two back-illuminated cameras operating in the B+V bands similar to \gaia\ $B_P$ (FLI Kepler KL400 sCMOS camera, $2048\times2048$ pixels), and in a red band close to the \gaia\ $R_P$ band (Andor iKon-L 936 CCD camera, $2048\times2048$ pixels). These cameras have an image scale of 1.05 and 1.30\,\arcsec\,pixel$^{-1}$ respectively resulting in $36\times36$ arcmin$^2$, and $44\times44$ arcmin$^2$ corrected fields of view \citep{Schmider_ASTEP_2022,Dransfield2022}. The fast full frame reading rate of the sCMOS sensor in the blue channel is used to guide the telescope mount at a typical rate of 0.5\,Hz, and these short exposure images are stacked to generate typical exposure times of about 1 minute. However, due to hardware signal transmission issues during the winter campaign of 2022, the blue camera could only be used during half of the season, thus leaving the red channel Andor camera as the only scientific detector for the rest of the austral winter. Exposure times and observation dates are given in Table\,\ref{tab:obslog}.

Due to the low data transmission rate at the Concordia Station, the data were processed on-site using an automated IDL-based pipeline described in \citet{2013A&A...553A..49A}. The calibrated light curve is reported via email and the raw light curves of about 1000 stars are transferred to Europe on a server in Rome, Italy, and are then available for deeper analysis. These data files contain each star’s flux computed through $10$ fixed circular aperture radii so that optimal light curves can be extracted. 

In total, we obtained 10 usable transits of HIP\,67522bc (five of each planet). Two additional transits of planet b were not usable due to insufficient baseline or low precision (poor weather). 

\subsection{Minerva-Australis}

We obtained two transits with telescopes at Mt Kent Observatory, Queensland, Australia. The first (full) transit was obtained using two 0.7\,m (T4 and T5) telescopes \citep{Addison2019_MinervaAustralis}, and the second (partial) was obtained with a single telescope (T4). Both telescopes are equipped with ZWO ASI6200MM PRO CMOS detectors, with no filter. Read out was done with 4x4 binning with a pixel scale of 0.5\arcsec/pixel.

We performed photometric extraction with \texttt{AstroImageJ} \citep{Collins2017}, which uses simple aperture photometry and sky-background subtraction. \texttt{AstroImageJ} automatically removes any comparison star trends by comparing the flux in its aperture to the sum of the flux in all other comparison star apertures. Comparison stars were selected and weighted to provide the highest-precision light curves.

\subsection{Swope}

We observed two transits of HIP\,67522\,c using the 1.0-m Henrietta Swope Telescope at the Las Campanas Observatory in Chile equipped with an E2V 4K$\times$4K CCD.  We observed on the local nights of 
2026\,Feb\,24 and 
2026\,Apr\,08. For both transits, we observed using a Sloan $r$ and neutral density filter to avoid saturation, and applied a mild defocus to improve photometric stability. We placed the target on quadrant 3 of the CCD to minimize non-linearity. On 
Feb\,24, we followed HIP\,67522 from UT 02:57--09:34, and on Apr\,08 from UT 23:45--09:55.  Conditions on both nights were clear, with similar $\approx$$2\farcs0$ seeing in the defocused images. The Apr\,08 night
had brief guiding problems around 08:30, which caused a data gap of a few minutes in duration.

We flat-fielded the images using sky flats taken during twilight, and reduced them to light curves using {\tt AstroImageJ} \citep[][]{Collins2017}.  We used a 12\,pixel ($5\farcs2$) radius circular aperture, with an outer annulus from $\approx$12$''$-17$''$ for background estimation.  We selected five nearby comparison stars, which by necessity were slightly fainter than the target, and found for both nights that the raw counts in the comparison star ensemble provided a good match to the target, irrespective of the exact choice of star.

\subsection{PEST}

We observed a single transit of HIP\,67522\,b on UTC 2026 April 13 in Sloan r' using the CDK14 telescope at the Perth Exoplanet Survey Telescope (PEST) observatory located near Perth, Australia.  The f/7.2 0.365 m telescope is equipped with a $6252\times4176$ QHY268M camera. Images are binned 2x2 in software giving an image scale of 0$\farcs$6 pixel$^{-1}$ resulting in a $32\arcmin\times21\arcmin$ field of view. A custom pipeline based on {\tt C-Munipack}\footnote{http://c-munipack.sourceforge.net} was used to calibrate the images and extract differential photometry using a circular $4\farcs8$ photometric aperture.

\section{Transit Fitting}\label{sec:transitfitting}

We (re-)analyzed transits of the two planets with the goal of keeping the analysis homogeneous (where possible) while also achieving the highest precision possible given the data. Simultaneous fitting of all transits would achieve both of these aims, as it naturally accounts for the strong covariances between transit parameters. However, each dataset is subject to a different set of systematics. Although the stellar variability can be fit with a single underlying model, instrumental signals from \tess, \cheops{}, \spitzer, and \jwst, differ from each other and from the atmospheric signals that frequently dominate ground-based data. Further, astrophysical variations (e.g, spot crossings and changes in transit depth from the planetary atmosphere) can lead to small but non-trivial changes in the transit observables. Lastly, a global model also proved to be computationally infeasible given the number of free parameters and total data volume involved.

As a result, we adopted a mixed strategy. All \tess\ data were fit simultaneously, and we used the outputs of this fit to put priors on the transit parameters for the remaining transits, which were each fit separately. The priors were sufficiently broad to allow for some flexibility from astrophysical variations, and we were able to adjust our systematics/variability model to the dataset at hand. 

The resulting observed and calculated times for all dataset are presented in Table\,\ref{tab:obslog}. 

\subsection{\tess\ transits and \texttt{Juliet}}\label{sec:juliet}

We fit the \tess\ data using \texttt{juliet} (Joint Analysis of Exoplanetary Transits \& RVs) \citep{espinoza2019juliet}. The \texttt{juliet} package utilizes \texttt{batman} \citep{Kreidberg2015} to model transits, \texttt{celerite} \citep{celerite} to model stellar variability, and \texttt{emcee} \citep{emcee} to explore parameter space. 

Using \texttt{juliet}, we simultaneously modeled the transit parameters of both planets, the transit timings of each individual transit of both planets, and the stellar variability. We used a double simple harmonic oscillator (SHO) Gaussian Process (GP) kernel following the \texttt{RotationTerm} kernel implementation in \texttt{celerite2} \citep{celerite2}. The GP kernel is described by the period ($P_{GP}$), the standard deviation of the process ($\sigma_{GP}$), the quality factor for the secondary oscillation ($Q_0$), the difference between the quality factors of the first and second mode ($\Delta Q$), and the fractional amplitude of the secondary mode compared to the primary ($f_{GP}$). We placed a Gaussian prior on $P_{GP}$ following the calculated rotation period of the star.

In addition to the GP parameters, there are three required instrumental parameters: $m_{\rm dilution}$ (dilution factor), $m_{\rm flux}$ (offset relative flux) and $\sigma$ (jitter term to account for underestimated uncertainties). Since there were no external sources of contamination, the dilution factor was fixed at 1. We found $\sigma$ was poorly constrained (approached 0), so we fixed it to 0.

For each planet, we fit for four unique transit parameters: the planet-to-star radius ratio ($R_p/R_*$), the impact parameter ($b$), $\sqrt{e}\sin{\omega}$, and $\sqrt{e}\cos{\omega}$ (to fit for the eccentricity, $e$, and the argument of periastron, $\omega$). \texttt{Juliet} would occasionally crash at high eccentricities, so we limited both parameters to the range of -0.7 to +0.7; more extreme values are unlikely for astrophysical (dynamical stability) reasons. We placed (uniform) priors on the other transit parameters following physical limitations. We also fit for a stellar density ($\rho$) and quadratic limb darkening parameters ($q_{1}$ and $q_{2}$), common to both planet models. We placed Gaussian priors on $\rho$ following the derived stellar parameters (0.45$\pm$0.04 $\rho_{\odot}$) and the limb darkening coefficients to account for differences in model predictions.

For speed, we binned the out-of-transit data to 10-minute bins, but preserved the fastest cadence for the window centered at the expected transit timings (assuming linear ephemerides) and two-times the expected transit duration. Tests on smaller subsets of the data and shorter chains indicated this did not impact the planet parameters; the only parameters being constrained by the out-of-transit data are the GP parameters, and the GP varies on timescales longer than 20-minutes. 

\begin{figure*}
    \centering
    \includegraphics[width=0.98\textwidth]{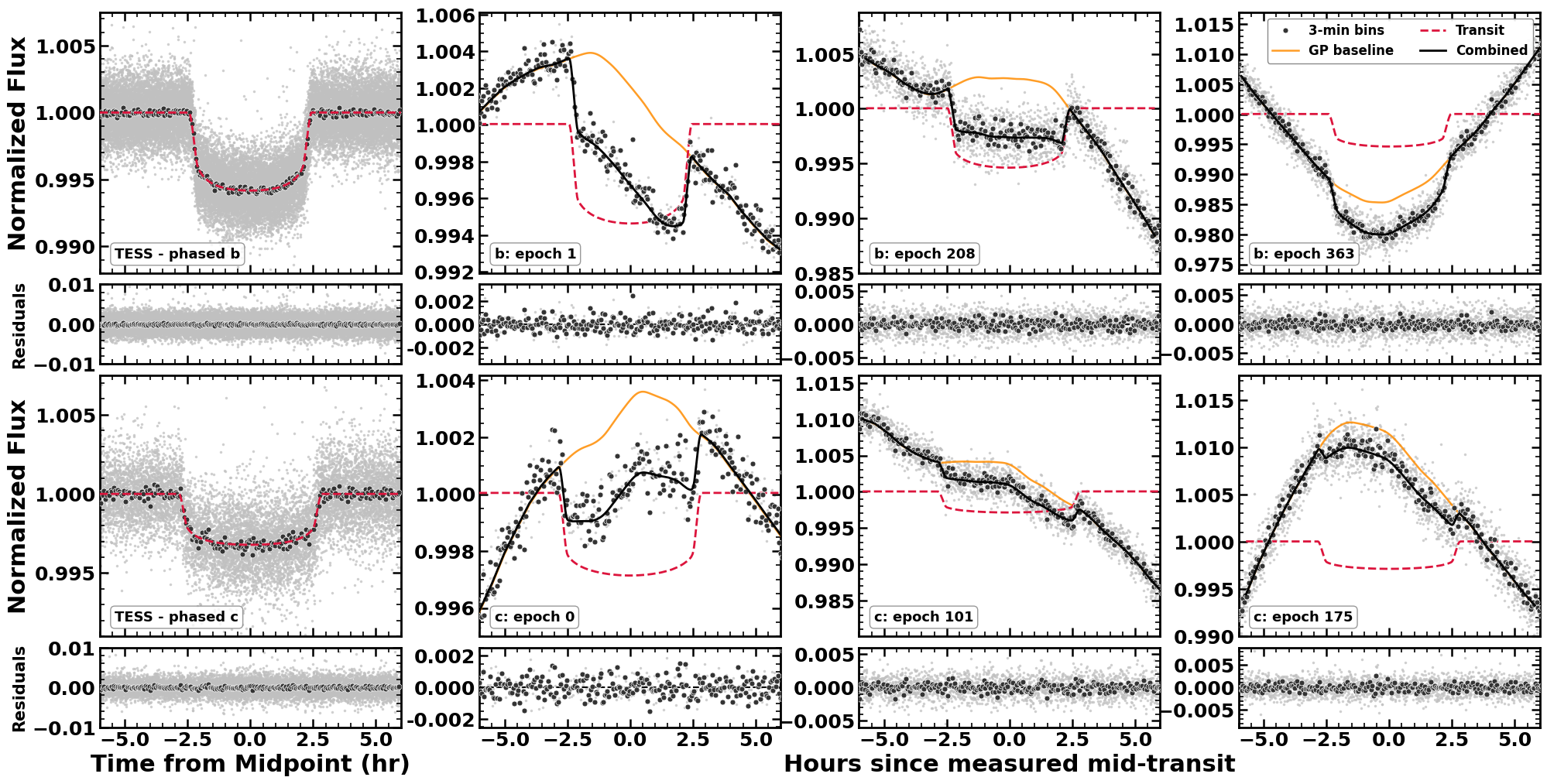}
    \caption{Left) Phase-folded \tess\ light curves (gray) highlighting transits of \plnameb\ (top) and \plnamec\ (bottom), binned to 3-minute intervals (black) for clarity. The GP variability model has been removed and the transit timings have been corrected for TTV offsets. The best-fit transit models are shown as the red-dashed lines. Right) Representative individual transits of \plnameb\ (top) and \plnamec\ (bottom). The stellar variability GP model is shown in orange, the transit model in red, and the combined model in black. The out-of-transit data in each panel is re-normalized to one for plotting clarity. The residuals of each data set are shown in the lower panels. }
    \label{fig:TESS_transits}
\end{figure*}

Overall, we fit for 45 parameters using \texttt{juliet}'s implementation of \texttt{dynesty}. We used 1000 live points. We list all parameters (except individual transit times), their priors, and their posteriors in Table\,\ref{tab:juliet}. We show a few representative transits and fits in Figure\,\ref{fig:TESS_transits}.

\begin{table*}
\centering
    \caption{\texttt{Juliet} posteriors of \plnameb\ and \plnamec}
    \begin{tabular}{lccc}
    \hline
    Description & Parameter & Prior$^{\alpha}$ & Posterior \\
    \hline
    \hline
    \multicolumn{4}{c}{Planet b}\\
    \hline
    planet-to-star radius ratio & $R_p/R_*$ & $U(0,1)$ & $0.0706\pm0.0007$\\
    impact parameter & $b$ & $U(0,1)$ & $0.091^{+0.064}_{-0.054}$ \\
    & $\sqrt{e}\sin{\omega}$ & $U(-0.7,0.7)$ & $-0.06^{+0.07}_{-0.08}$\\
    & $\sqrt{e}\cos{\omega}$ & $U(-0.7,0.7)$ & $-0.05^{+0.23}_{-0.20}$\\
    
    \hline
    \multicolumn{4}{c}{Planet c}\\
    \hline
    planet-to-star radius ratio & $R_p/R_*$ & $U(0,1)$ & $0.0526\pm0.0015$\\
    impact parameter & $b$ & $U(0,1)$ & $0.369^{+0.064}_{-0.114}$\\
    & $\sqrt{e}\sin{\omega}$ & $U(-0.7,0.7)$ & $-0.03^{+0.12}_{-0.10}$\\
    & $\sqrt{e}\cos{\omega}$ & $U(-0.7,0.7)$ & $0.18^{+0.28}_{-0.31}$\\
    
    \hline
    \hline
    \multicolumn{4}{c}{Global Parameters}\\
    \hline
    stellar density & $\rho_*$ ($\rho_\odot$) & $N(0.45,0.04)$ & $0.449^{+0.026}_{-0.015}$ \\ 
    limb-darkening parameter & $q_1$ & $N(0.335,0.04)$ & $0.275^{+0.027}_{-0.026}$\\
    limb-darkening parameter & $q_2$ & $N(0.368,0.03)$ & $0.353^{+0.025}_{-0.024}$\\
    GP period & $P_{GP}$ & $N(1.42,0.02)$ & $1.4252^{+0.0027}_{-0.0028}$\\
    GP standard deviation & $\sigma_{GP}$ & $LU(10^{-5},0.1)$ & $0.0122\pm0.0007$\\
    GP quality factor & $Q_0$ & $LU(0.1,500)$ & $0.553^{+0.044}_{-0.034}$\\
    GP difference between quality factors & $\Delta Q$ & $U(0,500)$ & $91^{+6}_{-9}$\\
    GP fractional amplitude & $f_{GP}$ & $U(0,1)$ & $0.275^{+0.051}_{-0.042}$\\
    relative flux offset & m$_{flux}$ & $N(0,0.1)$ & $-0.00008\pm0.00020$\\
    \hline
    \hline
    \multicolumn{4}{l}{$^{\alpha}$ $U(a,b)$ indicates a uniform prior from $a$ to $b$.}\\
    \multicolumn{4}{l}{\hspace{0.15cm} $N(a,b)$ indicates a normal prior centered at $a$ with standard deviation $b$.}\\
    \multicolumn{4}{l}{\hspace{0.15cm} $LU(a,b)$ indicates a log-uniform prior from $a$ to $b$.}\\
   \multicolumn{4}{l}{The individual transit times were also simultaneously fit. We denote their posteriors in Table\,\ref{tab:obslog}.}
    \end{tabular}
    \label{tab:juliet}
\end{table*}

\subsection{CHEOPS}\label{subsec:transit_cheops}

\cheops\ photometry exhibits instrumental (systematic) flux variations correlated with the spacecraft's roll angle and background sky level, which must be fitted simultaneously with the transit \citep[e.g.,][]{Oddo2023}. Our mean model consists of a physical transit model generated via \texttt{batman} \citep{Kreidberg2015} multiplied by a linear detrending matrix. To capture the spacecraft systematics, the linear baseline model includes a second-order polynomial in time, the background flux, and the first three harmonics of the spacecraft roll angle ($\phi$): $\sin(\phi)$, $\cos(\phi)$, $\sin(2\phi)$, $\cos(2\phi)$, $\sin(3\phi)$, and $\cos(3\phi)$. To account for residual correlated noise driven by stellar variability and low-level instrumental jitter, we included a Gaussian Process using \texttt{celerite2} \citep{celerite} with a Mat\'ern-3/2 kernel. We also included a white-noise jitter term to accommodate underestimated photometric uncertainties. 

For each visit, we ran the fitting procedure in two stages. First, we performed an initial quick-fit using a least-squares linear solver combined with the expected planetary transit model. We computed the residuals from this initial model and performed a robust $5\sigma$ clip based on the median absolute deviation (MAD). We then ran a longer fit with Markov Chain Monte Carlo (MCMC) described below. Upon inspecting the initial MCMC outputs, we manually identified two large post-transit flares and a small number (less than 20 points per observation) of individual outlier points. The outliers were manually flagged and removed. The flares were fit using an exponential flare model, which includes an amplitude, a start time, and a decay timescale. 

We placed Gaussian priors on the transit model (e.g., $R_p/R_*$, impact parameter $b$, and stellar density $\rho_*$) drawn from our \texttt{Juliet} fit. We fixed limb-darkening coefficients to values from \texttt{LDTK} \citep{Parviainen2015}. We locked $P$ to the \texttt{Juliet} result, but let $T_0$ evolve under a uniform prior bounded only by the limits of the observations. We assumed zero eccentricity. For the four full transits, we used uniform priors on the three GP hyperparameters (the amplitude $\sigma_{\rm{GP}}$, timescale $\rho$, and jitter $f$). For $\rho$, we limited it to $>0.1$\,days, because shorter timescales overlap with the instrumental signal and we want that handled by the matrix above. 

The partial transits were more complicated; without sufficient out-of-transit baseline on both sides of the transit event, the unconstrained GP was excessively flexible and often unphysically altered the transit shape to minimize the local residuals. While no fit failed because of this (they all converged), many fits had walkers wandering to obviously erroneous solutions (e.g., removing the full transit with the GP and instead fitting a data gap as the egress). These bad fits always corresponded to high values of $\sigma_{\rm{GP}}$ and usually smaller values $\rho$ (high amplitude and short timescale) when compared to the full transits. To prevent this, we derived priors for the GP hyperparameters ($\sigma$ and $\rho$) from the spread in our four full transit fits and applied these as (weak) Gaussian priors.

Finally, one visit (Epoch 154 for planet c, corresponding to Epoch 318 for planet b) contained an overlapping partial transit of both HIP\,67522\,b and HIP\,67522\,c. For this epoch, we expanded the dimensionality of our MCMC to model both planets simultaneously, combining their respective \texttt{batman} models and linking them via a shared stellar density parameter to ensure physical consistency. 

\begin{figure*}[htb]
    \centering
    \includegraphics[width=0.48\textwidth]{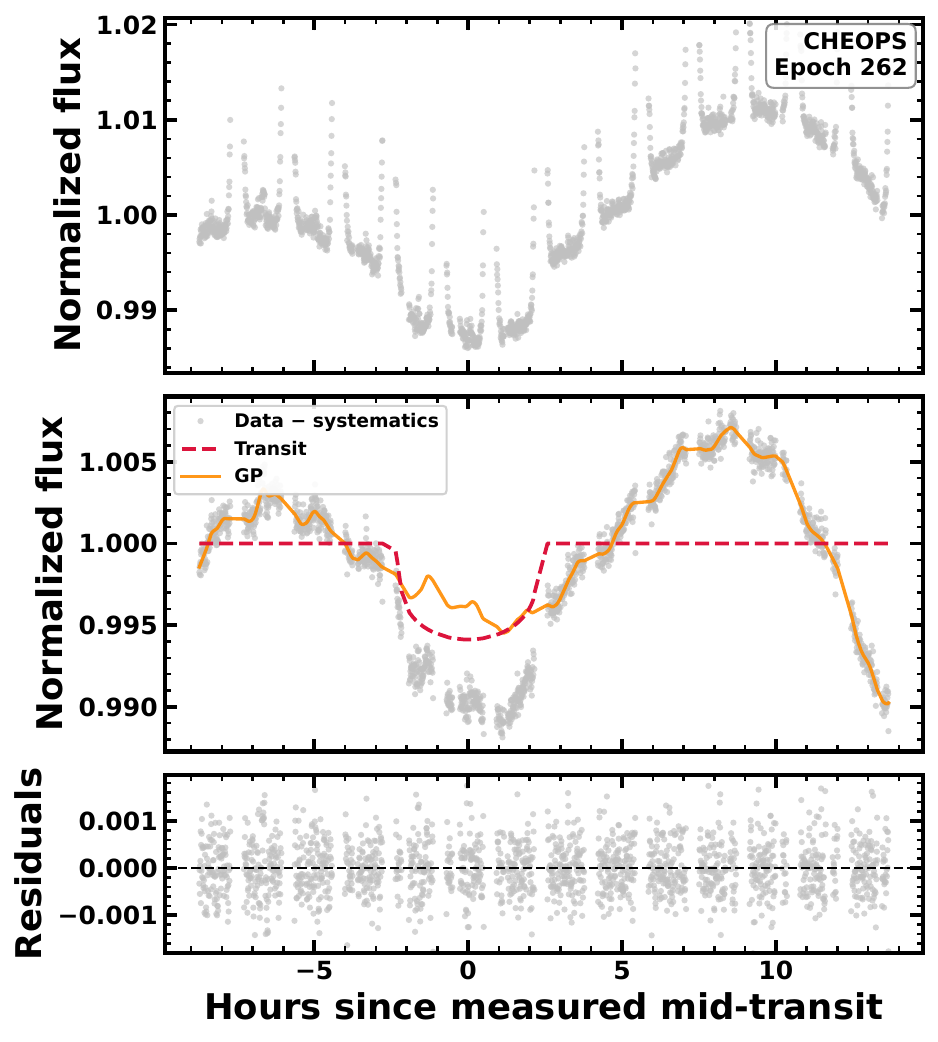}
    \includegraphics[width=0.48\textwidth]{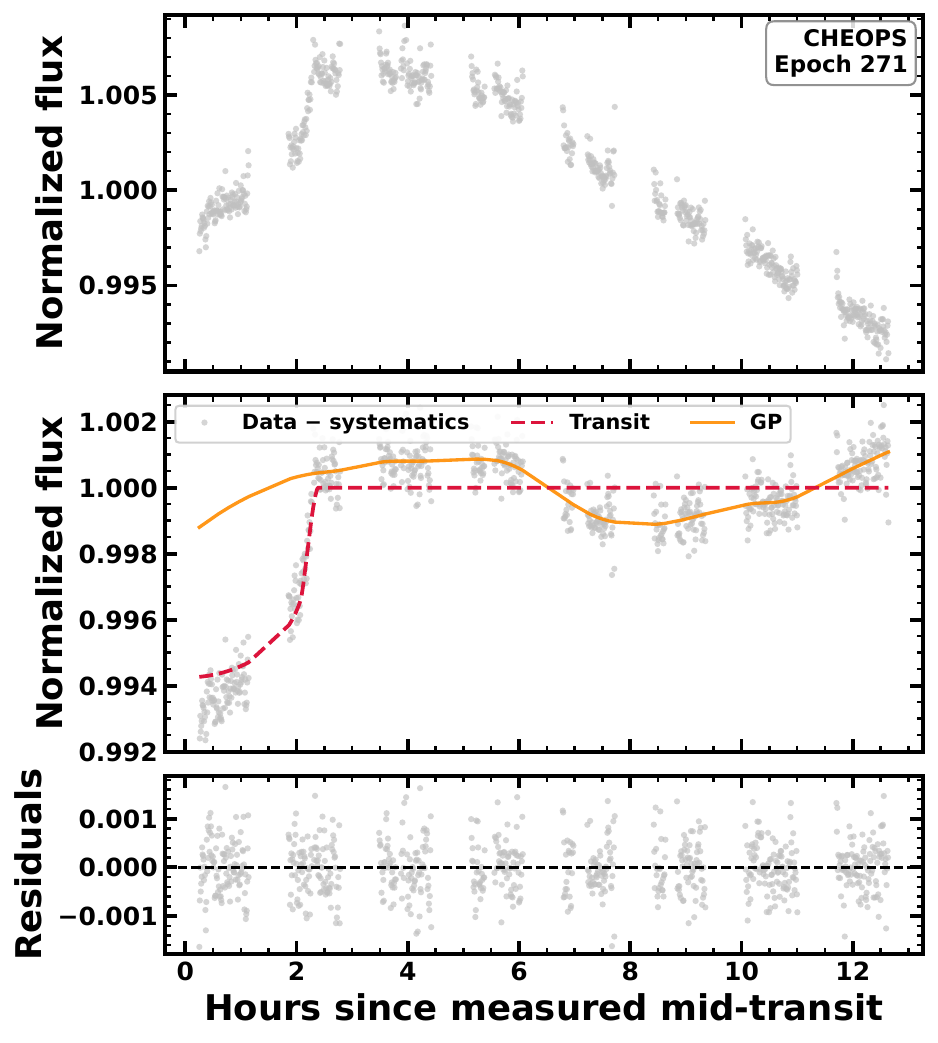}
    \caption{Representative \cheops{} transits of HIP\,67522b. Left shows a full transit, while right is one of the partials. The top panel displays the raw flux of the transit. The middle panel has the systematics model subtracted out and GP variability and transit model shown (orange and red lines). Note that instrumental systematics, stellar variability, and transit parameters are all fit simultaneously. The bottom is the final residual. Fits are well-behaved and systematics are properly removed with the combined model, enabling timing precision of 1-5\,minutes for \cheops{} transits. }
    \label{fig:cheops_full}
\end{figure*}

Examples of the detrending and fitting results are shown in Figure\,\ref{fig:cheops_full} for representative full and partial transits. 

\subsection{Ground-based data (LCOGT, SOAR, MINERVA-Australis, Swope, PEST, and ASTEP)}

As with \cheops, we fit epoch individually (simultaneous data from the same facility were analyzed together), fitting the transit with a \texttt{batman} model and the variability (stellar or atmospheric) using the Mat\'ern-3/2 kernel from \texttt{Celerite}, all within a MCMC framework using \texttt{emcee} and using the same priors on the transit parameters. We also tested 1) a simple linear model in airmass, 2) an exponential model in airmass, and 3) a linear airmass model with a linear time correction. We found these to be almost entirely degenerate with the GP, likely because airmass varies smoothly and on timescales comparable to the observational window. Turning off the GP and just using an airmass and time-based model often yielded more precise timings, but also left significant structure in the residuals. That approach likely leaves us vulnerable to spot-induced timing variations \citep{Mazeh2015a,LopezMurillo2026}. 

For each filter, we locked the quadratic limb-darkening parameters to the values from \texttt{LDTK}, listed in Table\,\ref{tab:ld_priors}. The data is not precise enough to provide additional constraints on the limb-darkening coefficients and the GP adjusts to variations in limb-darkening. The remainder of the transit parameters are the same as the \cheops\ fits.

\begin{figure}
    \centering
    \includegraphics[width=0.48\textwidth]{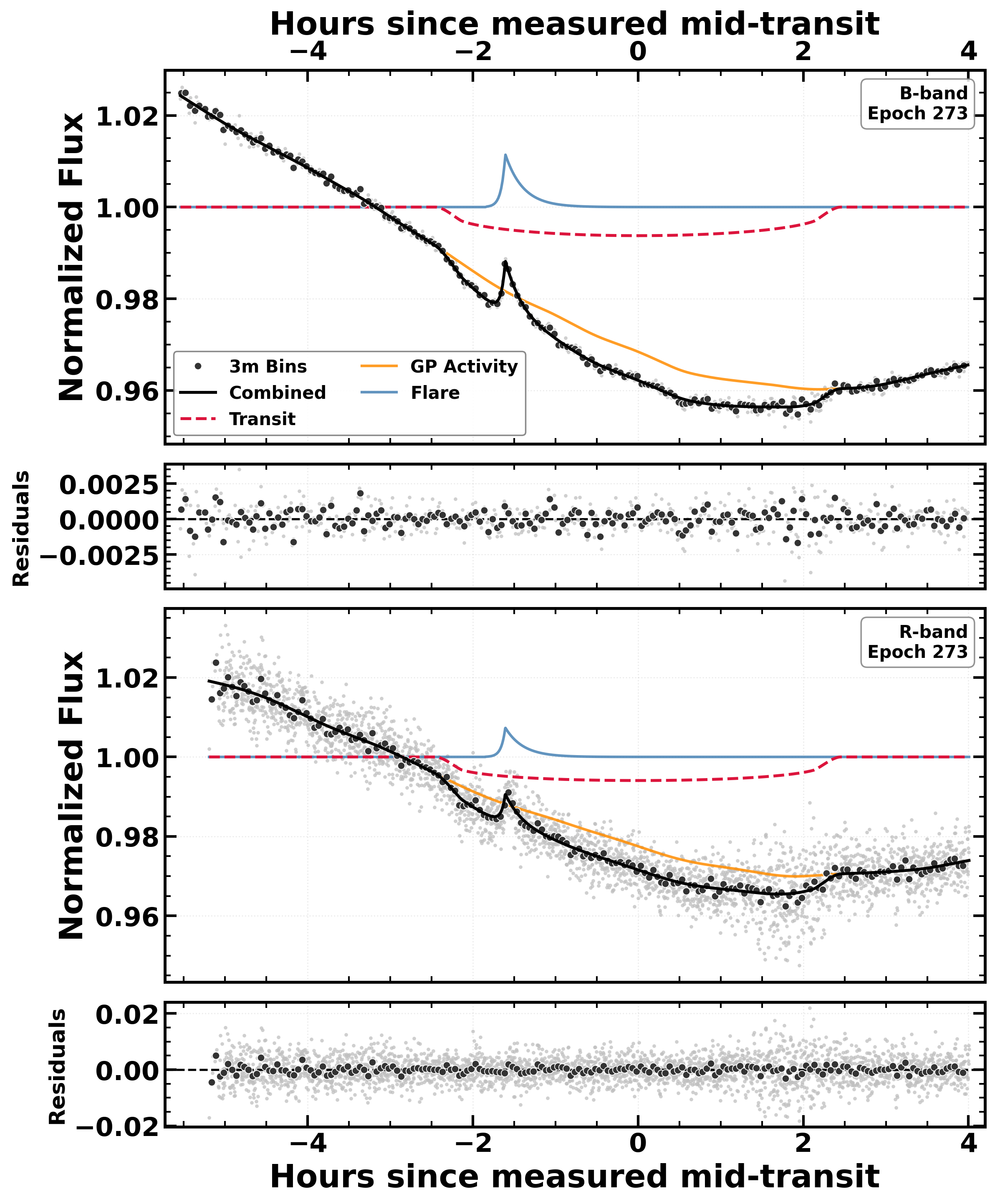}
    \caption{Representative ASTEP transit of HIP\,67522\,b. The top panel shows the $B$-band data and the bottom the $R$-band data. For clarity, we show 3-min bins of the photometry. These data have a clear flare early in the transit, which we model with a simple exponential (blue). The stellar variability and atmospheric variations are handled by a GP kernel (orange). This preserves the transit model (red). The combined model is in black, with the residuals shown in the bottom panels of each dataset. The combined model is excellent, as is evident in the lack of correlated noise in the residuals. }
    \label{fig:astep}
\end{figure}

We manually flagged clear outliers and flares; flares were fit with an exponential model (Figure\,\ref{fig:astep}) while outliers were manually removed. Inspection of a subset of removed points suggests most were due to problems with the observations themselves (e.g., wind shakes, large guiding corrections, or cosmic rays). Fewer than 0.1\% of points were removed this way.

We ran all of the above transit fits using 32 walkers for at least 5000 steps. After the initial 5000, the MCMC chain was checked every 5000 steps for convergence, requiring at least 50$\times$ the autocorrelation time to stop the run. The longest chains took 60,000 steps to converge.

There was some variation in the fitting procedure based on the instrument, transit coverage, number of wavelengths, and presence of data gaps. We provide specific details on these cases below:

{\bf ASTEP:} The ASTEP data contain data across two filters ($B$ and $R$). For two transits (2024\,Jul\,31 and 2024\,Aug\,29) the $B$ data were too low quality to be used (high scatter and large data gaps). For the rest, the fit was done on both datasets simultaneously with a common GP timescale ($\tau_{\rm{GP}}$), but different amplitudes ($\sigma_{\rm{GP}}$), limb-darkening parameters, and jitter terms. The assumption is that the dominant source of variation is from the star and Earth's atmosphere, which have the same timescales between filters but differ in amplitude. For light curves with flares, the flare location and decay timescale were forced to match between observations, but the amplitude was allowed to vary (adding an additional free parameter). An example of this is shown in Figure\,\ref{fig:astep}.

\begin{figure}[htb]
    \centering
    \includegraphics[width=0.48\textwidth]{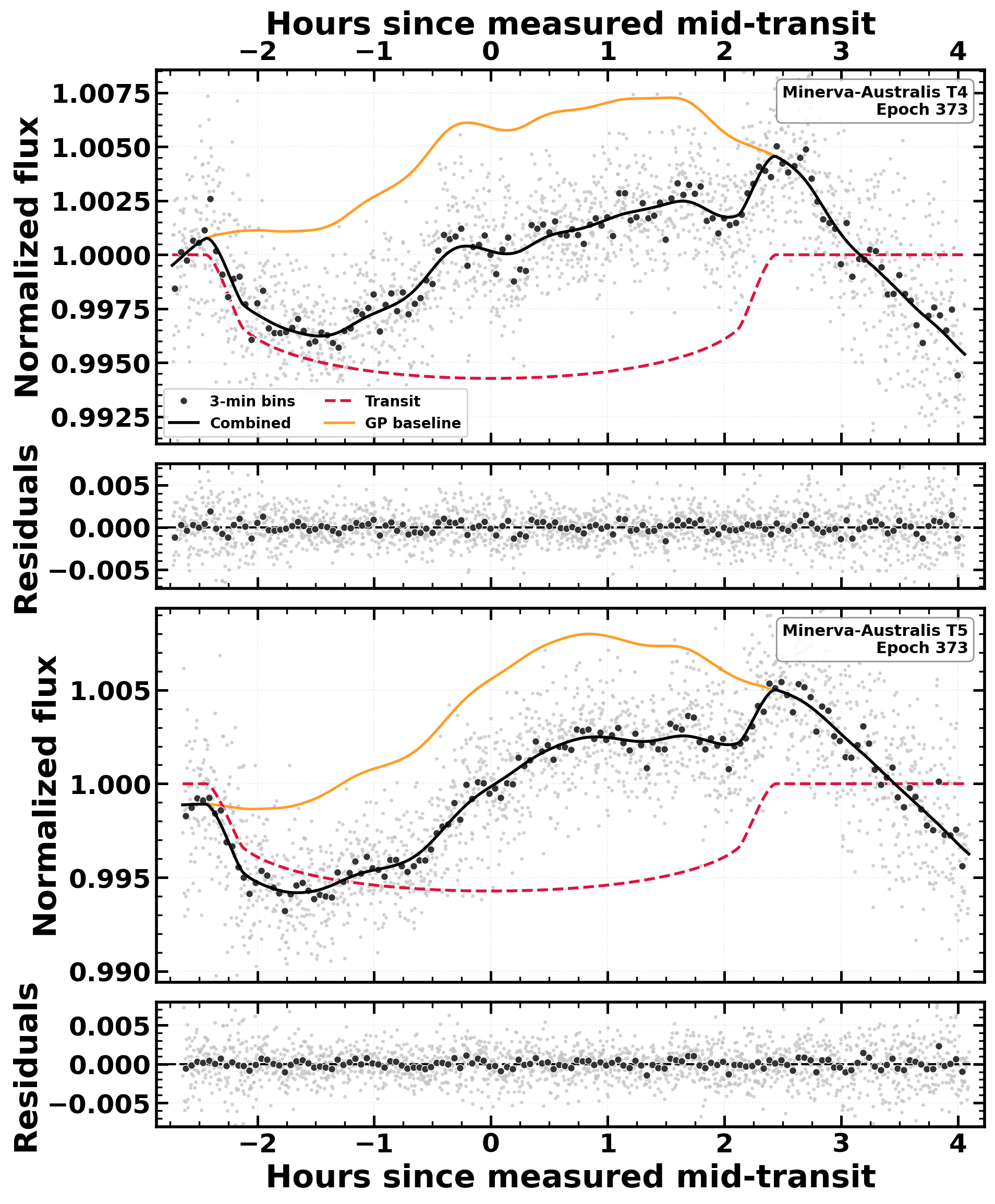}
    \caption{Same as Figure\,\ref{fig:astep}, but for observations of HIP\,67522\,b taken with two of the MINERVA-Australis telescopes (T4 and T5). The two transits show similar, but not identical GP variability, confirming that the GP is handling both stellar and instrumental variability.  }
    \label{fig:minerva}
\end{figure}

{\bf MINERVA-Australis:} One transit observed from MINERVA-Australis was taken with two telescopes simultaneously (the second partial transit was covered by only one telescope) using a single common filter. As with ASTEP, the fit was done on both datasets simultaneously with a common GP timescale but letting the amplitude vary. The transits are observed through the same column of atmosphere and of the same star; hence the GP amplitudes should be the same. However, as we show in Figure\,\ref{fig:minerva}, some of the systematics vary between the two instruments, perhaps due to instrumental signals or small variations in the reduction. Ultimately, untying the GP amplitude between telescopes did not impact the transit time constraints but produced less structured residuals. 

{\bf Simultaneous LCOGT transits:} Many transits were observed by two different facilities simultaneously. This happens when multiple teams were pursuing the same target or because of multiple LCOGT requests (to increase the likelihood of success). In such cases, they were analyzed together in the same way as the MINERVA-Australis data.

\begin{figure}[htb]
    \centering
    \includegraphics[width=0.49\textwidth]{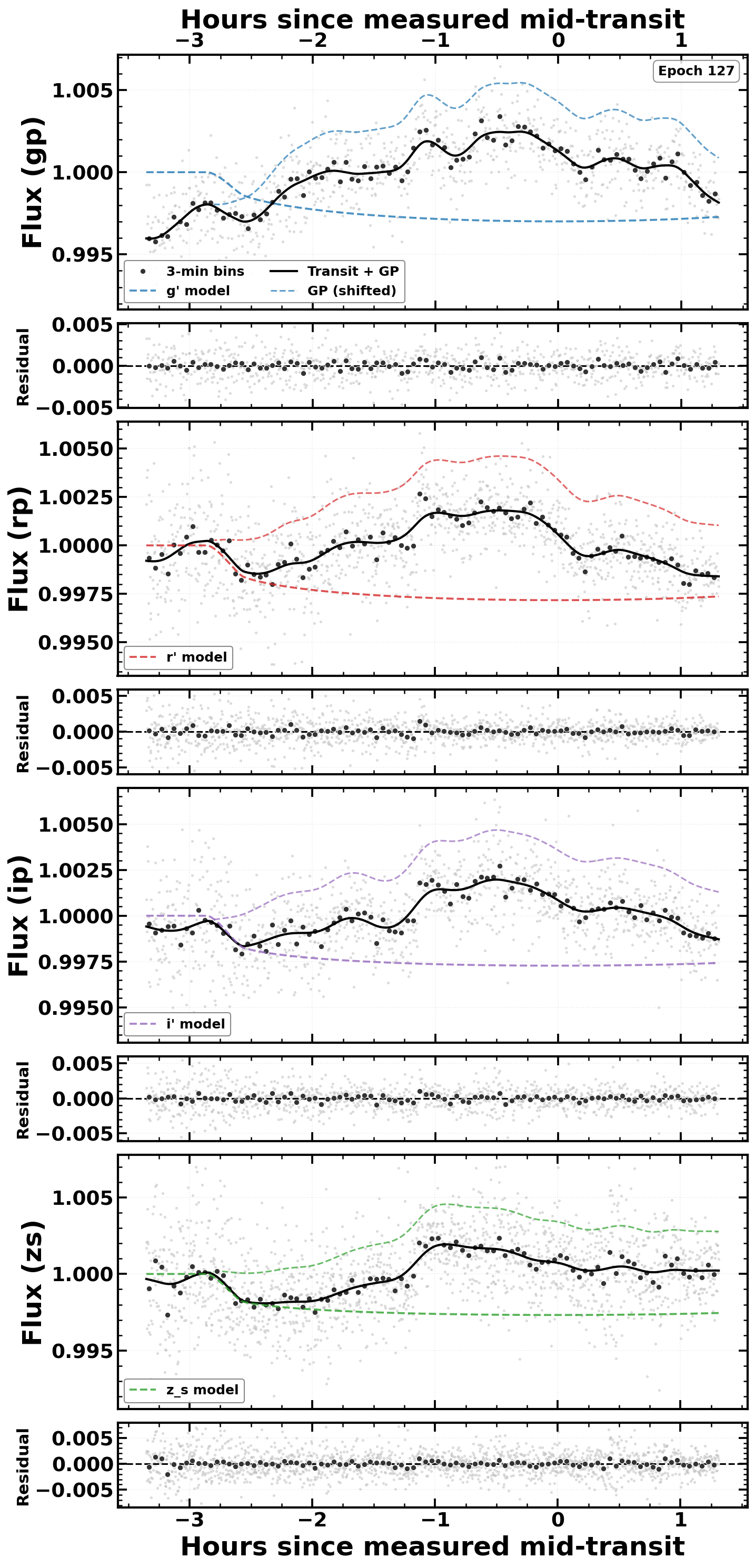}
    \caption{Same as Figure\,\ref{fig:astep}, but for observations of HIP\,67522\,c taken with the 4-channel MuSCAT4 imager ($g, r, i, z_s$). The data shows systematics that are common across all four filters, but vary slightly in amplitude and shape. These are likely from the star (stellar variations, flares, or spot crossings), and therefore show chromaticity (most structure in $g'$ and least in $z_s$). The GP parameters are well-constrained by the joint fit, and the complex structure is well-modeled by the GP kernel. }
    \label{fig:muscat4}
\end{figure}

{\bf MuSCAT4:} MuSCAT4 simultaneously observes the transit in four filters, each with their own exposure times. As with the above simultaneous observations, we fit these using a common GP timescale but allow the jitter term and amplitude to vary. We show an example in Figure~\ref{fig:muscat4}.

{\bf Partials:} For many partial transits, we found that our initial fits yielded extremely poor transit time (errors of up to 10\,minutes). This was, in part, because we use two (independent) Gaussian priors on $\rho$ and $b$, rather than simultaneously fitting the data as was done for all \tess\ transits. The MCMC could therefore explore transit durations that were inconsistent with the \tess\ data (and all other data) by exploring ($\rho,b$) sets that would have been excluded by a joint fit. We tested using a two-dimensional prior built using kernel density estimation from the \texttt{Juliet} posterior. This reduced but did not eliminate the issue. Instead, we found more physically consistent results by placing a modest prior on the transit duration ($\pm$0.15\,hours). 

We used four full transits from LCOGT and ASTEP (two from each) to test if this duration prior was biasing our results or uncertainties. For this test, we trimmed out either ingress or egress (randomly), and refit the data. We repeated this 100 times, and found that the resulting times matched the full-transit fits perfectly (median difference of $<1$\,min) and the scatter in measured transit times was {\it smaller} than the uncertainties on the fits (RMS/$\sigma_{\rm{Tc}}\simeq0.7$). This suggests we are overestimating our errors, likely because of the lack of baseline to tune the GP parameters.

{\bf Data gaps or jumps:} A few ground-based datasets showed `jumps' in the light curve from guiding losses (e.g., weather or technical issues). The first two SOAR transits, for example, had large data gaps due to the temporary SOAR elevation limits (Section\,\ref{sec:SOAR}). In such cases, the target and comparison stars usually land on a slightly different set of pixels before and after the jump. As with flares, these were manually identified and corrected by applying a small shift in the data estimated from a global least-squares fit to the data. Except for the second SOAR transit (Figure\,\ref{fig:soar}; which we excluded from our analysis) none of the jumps occurred close to ingress or egress, so these did not impact the timing measurement significantly.

\begin{deluxetable}{lrrr} 
\tabletypesize{\footnotesize} 
\tablecaption{Limb-darkening values} \label{tab:ld_priors}
\tablewidth{0pt}
\tablehead{
  \colhead{Filter} &
  \colhead{$\lambda_0$ (nm)} &
  \colhead{$u_1$} &
  \colhead{$u_2$}
}
\startdata
  TESS            &   798.8 &  0.425 &  0.154 \\
  MINERVA-Australis &  640.0 & 0.541 & 0.132\\
  CHEOPS          &   641.2 &  0.550 &  0.129 \\
  ASTEP $B$         &   544.5 &  0.628 &  0.111 \\
  ASTEP $R$         &   850.0 &  0.388 &  0.154 \\
  $g'$              &   478.2 &  0.729 &  0.067 \\
  Narrow $g$      &   515.0 &  0.665 &  0.100 \\
  $r'$             &   626.0 &  0.527 &  0.146 \\
  Na D           &   589.0 &  0.563 &  0.138 \\
  $i'$              &   773.2 &  0.420 &  0.154 \\
  Narrow $i$      &   792.0 &  0.408 &  0.154 \\
  $z_s$              &   867.9 &  0.366 &  0.156 \\
  Narrow $z$        &   868.0 &  0.362 &  0.156 \\
  \spitzer      &  4470.7 &  0.105 &  0.079 \\
  NRS1        &  3323.5 &  0.130 &  0.108 \\
  NRS2        &  4447.0 &  0.107 &  0.081 \\
\enddata
\end{deluxetable}

\begin{figure}
    \centering
    \includegraphics[width=0.48\textwidth]{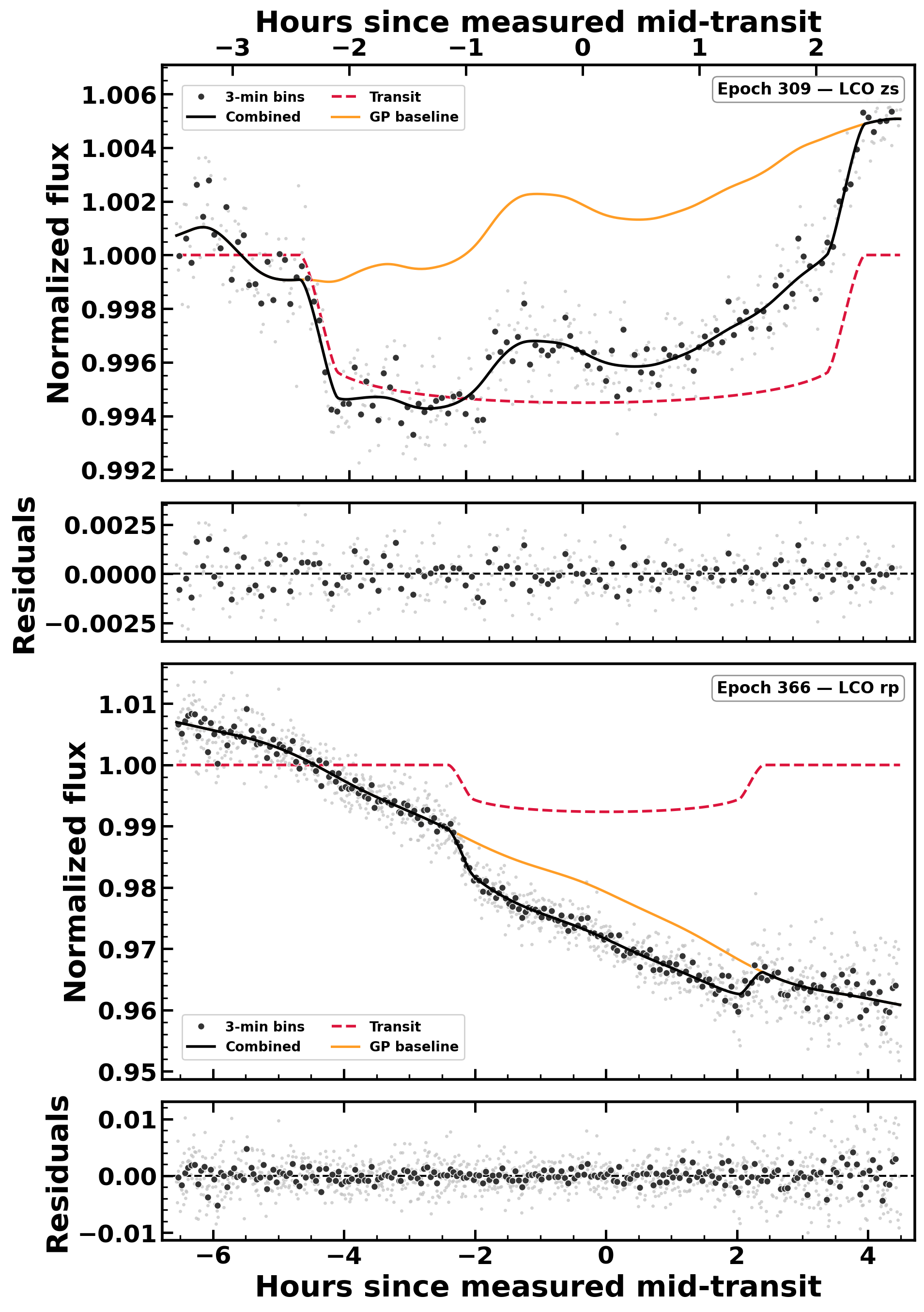}
    \caption{Same as Figure\,\ref{fig:astep}, but for two different LCOGT observations taken with different filters. A likely spot crossing is seen in the $z_s$ transit (top), but the Matern3/2 GP models it effectively, leaving behind little to no structure in the residual. The timings were unaffected. Modeling these with a polynomial for the out-of-transit trends and a Gaussian at the spot crossing yielded similar times, as seen in prior studies \citep{LopezMurillo2026}. }
    \label{fig:lcogt}
\end{figure}

\subsection{NGTS transit times}\label{subsec:ngts_data}

\citet{Chakraborty2026} presented three transits of HIP\,67522\,b taken with the Next-Generation Transit Survey \citep[NGTS;][]{Wheatley2018}. Since our transit fits and times were broadly consistent with those in \citet{Chakraborty2026} (see Section\,\ref{sec:summary}), and because the NGTS data are not public, we opted to take their transit times and uncertainties as part of our analysis. 

\section{Modeling the Transit Timing Variations}\label{sec:ttv}

Having measured the mid-transit times for each transit individually (Section\,\ref{sec:transitfitting}), we fit them jointly with a dynamical $N$-body model to constrain the planet masses and eccentricities. The transit-time posteriors are well-approximated as Gaussian for most transits, with a few showing mildly asymmetric tails. We collapsed each posterior to a single value, taking the median as the central value and the half-difference between the 84th and 16th percentiles as the symmetric standard deviation; in the few cases where the distribution was strongly asymmetric, we adopted the larger of the two sides as a conservative uncertainty. The resulting dataset comprises \btransit{} transits of planet b and \ctransit{} of planet c, listed in Table\,\ref{tab:obslog}.

For the TTV analysis, we adopted the following fixed parameters. The stellar mass was set to $M_\star = 1.22\,M_\odot$ \citep{THYMEII}. Locking the stellar mass rather than allowing it to float with a Gaussian prior leads to an underestimation of the planet masses in the TTV analysis, but this is negligible in comparison to the other sources of uncertainties. We used planet inclinations of $i_b = 89.88^\circ$ and $i_c = 89.20^\circ$ from our \texttt{Juliet} fit, and the longitude of ascending node was fixed at $0^\circ$ for both planets. For nearly coplanar systems, small changes in inclination produce negligible TTV signal changes \citep{Nesvorny2014}, and the inclinations are well-constrained by the transits. RM data also strongly suggests the system is well-aligned \citep{Heitzmann2021}. We ran some simple least-squares fits modifying these parameters within measurement uncertainties and found the changes were small compared to the formal uncertainties.

\newcommand{\ttvcode}{\texttt{TTVFast}}

\subsection{Forward model and likelihood}\label{sec:ttv_model}

Planets near first-order mean-motion resonances exhibit large-amplitude, characteristically sinusoidal TTVs whose period and amplitude depend on the proximity to resonance and the planet masses \citep{Lithwick2012}. We model these signals dynamically using \ttvcode{} \citep{Deck2014}, a fast symplectic $N$-body integrator optimized for transit timing problems. \ttvcode{} advances the system from a reference time $t_{\rm ref}$ given the masses and osculating Keplerian elements (period $P$, eccentricity $e$, argument of pericenter $\omega$, mean anomaly at $t_{\rm ref}$, inclination, longitude of ascending node) and outputs the predicted mid-transit times for each planet across the integration window. We set the step size to $0.02$ days ($\approx P_b/350$), $t_{\rm ref} = 1500$ TBJD (well before the first observed transit), and integrate forward to TBJD $4500$ (long after the last transit).

For each MCMC sample we run the integrator and pair each observed transit to  the nearest predicted model transit in time. The log-likelihood is the standard $\chi^2$:
\begin{equation}
\ln\mathcal{L} = -\frac{1}{2} \sum_i \frac{(T_{i,{\rm obs}} - T_{i,{\rm model}})^2}{\sigma_i^2},
\end{equation}
summed over all transits of both planets. Observational uncertainties ($\sigma_i$) are fixed per-transit values from our individual light-curve fits, so the Gaussian normalization is a constant and is omitted from $\ln{\mathcal{L}}$.

We explored four different fits (described below). For all fits, the orbital periods, times of first transit, and mass of the outer planet ($M_c$) evolve under uniform priors with generous outer bounds ($\pm$0.1\,days for $P$, $T_0$ and 1-50$M_\oplus$ for $M_c$). The other free parameters, eccentricity ($e_b$ and $e_c$), argument of periastron ($\omega_b$ and $\omega_c$), and the b-planet's mass ($M_b$) are handled differently in each fit:
\begin{enumerate}
    \item \textbf{Circular, informative mass prior.} 
    We fix $e_b = e_c = 0$ and adopt the \citet{Thao2024_featherweight} mass measurement as a Gaussian prior on planet b: $M_b = 13.8 \pm 1.0\,M_\oplus$. The free parameters are $(M_b, M_c, P_b, t_{0,b}, P_c, t_{0,c})$.
    
    \item \textbf{Eccentric, informative mass prior.}
    We allow eccentricity to 
    vary by adding $\sqrt{e_b}\cos{\omega_b}, \sqrt{e_b}\sin{\omega_b}, \sqrt{e_c}\cos{\omega_c}, \sqrt{e_c}\sin{\omega_c}$, for a total of 10 free parameters. \ttvcode{} asks for the traditional $(e, \omega)$ parameterization, but this is singular at $e=0$ (where $\omega$ is undefined), and creates a positive $e$ bias \citep[e.g.,][]{Wang2011}. So we retain the parameterization from the \texttt{Juliet} fit and perform the conversion inside the likelihood function before feeding into \ttvcode. Consecutive transits and the overall TTV shape provide some constraints on $e$ and $\omega$, but we improve on these by providing additional constraints from the transit-duration and stellar-density information \citep{Seager:2003lr}. We use the joint posterior on these parameters from our \texttt{Juliet} fit to build a kernel density estimate (KDE) prior (using \texttt{scipy.stats.gaussian\_kde}). For this fit, we retain the Gaussian prior on $M_b = 13.8 \pm 1.0\,M_\oplus$.
    
    \item \textbf{Circular, uninformative mass prior.} We repeat Fit\,1 with a uniform prior on $M_b$ between 1 and 50 $M_\oplus$. This serves as a test of the input mass prior for the b planet. 
    
    \item \textbf{Eccentric, uninformative mass prior.} This is the least restrictive configuration, combining the free $(e, \omega)$ parameters and KDE prior from Fit\,2 with the uniform $M_b$ prior as in Fit\,3. This explores what results we can get with a narrow set of assumptions and can be used to see if any of these assumptions are disfavored or ruled out by the data. 
\end{enumerate}

To initialize each MCMC fit, we first perform a Nelder-Mead least-squares optimization under the same likelihood and priors, which locates the maximum a posteriori (MAP) point; we then initialize walkers around the best-fit solution by adding a random draw from the Nelder-Mead uncertainties.

We then performed a full fit using \texttt{emcee} with 64 walkers. The eccentric fits (see below) were slow to converge using default \texttt{emcee} setup, so we use a mixture of DEMove (60\%), DESnookerMove (20\%), and StretchMove with $a=1.5$ (20\%). We found this mix to be substantially more efficient than the default stretch move alone. We check convergence every 5000 steps and stop when the chain length exceeds $50\tau$ and $\tau$ is stable to within 10\%, where $\tau$ is the maximum integrated autocorrelation time across all parameters. We discard the first $2\tau$ steps as burn-in. The two eccentric fits took more than 100,000 steps before they were well mixed.

\section{Results}\label{sec:results}

\subsection{Transit fit results}

Our transit fits yield relatively precise times, especially given the stellar variability and visible spot crossings. Transits from \tess{} and \spitzer{} achieve sub-minute precision. The least precise fits provide timing uncertainties of 4-6\,m, which come mostly from ground-based data (and the one \cheops{} visit with transits of both b and c). Timing constraints are also generally more precise for planet b than for c, due to a combination of a shallower transit, more partials (longer period and duration makes full transits a challenge), and a higher rate of spot crossings seen in the c data. 

Spot crossings are evident in multiple transits, as noted in prior studies \citep[e.g.,][]{Barber2024a,Thao2024_featherweight}. Fortunately, none of these appear to land on ingress or egress where they are at higher risk of inducing a transit time bias. Further, \citet{LopezMurillo2026} showed that GP regression, like was used here, naturally handles spot crossings, removing the bias (at the cost of higher uncertainties). Indeed, we can see the GP fitting out clear crossings in some transits (e.g. Figure\,\ref{fig:cheops_full}a, Figure\,\ref{fig:lcogt}a). 

We provide an updated (linear) ephemeris based on a simple least-squares fit to all transit timings in Table\,\ref{tab:ephemeris}. Table\,\ref{tab:juliet}, which has the results of the \texttt{Juliet} fit, provides the most precise values for other parameters (e.g., $R_P/R_*$). Though our fit prefers a slightly larger radius for planet b, $10.63\pm0.47$R$_\oplus$, the planetary parameters for planet b and c agree with previously reported values \citep{THYMEII,Heitzmann2021,Barber2024a}.

\begin{deluxetable}{lccc} 
\label{tab:ephemeris}
\tabletypesize{\footnotesize} 
\tablecaption{Effective linear ephemeris}
\tablewidth{0pt}
\tablehead{
\colhead{Planet} &
\colhead{$T_{0}$ (BJD)} &
\colhead{$P$ (days)} 
}
\startdata
b & 2458597.06435$\pm$0.00028 & $6.9594703^{+0.00000094}_{-0.0000010}$ \\ 
c & $2458602.50131^{+0.00075}_{-0.00081}$ & $14.3349281^{+0.0000061}_{-0.0000064}$ \\ 
\enddata
\end{deluxetable}

\subsection{TTV results}

\begin{figure*}[htp]
    \centering
    \includegraphics[width=1.05\textwidth]{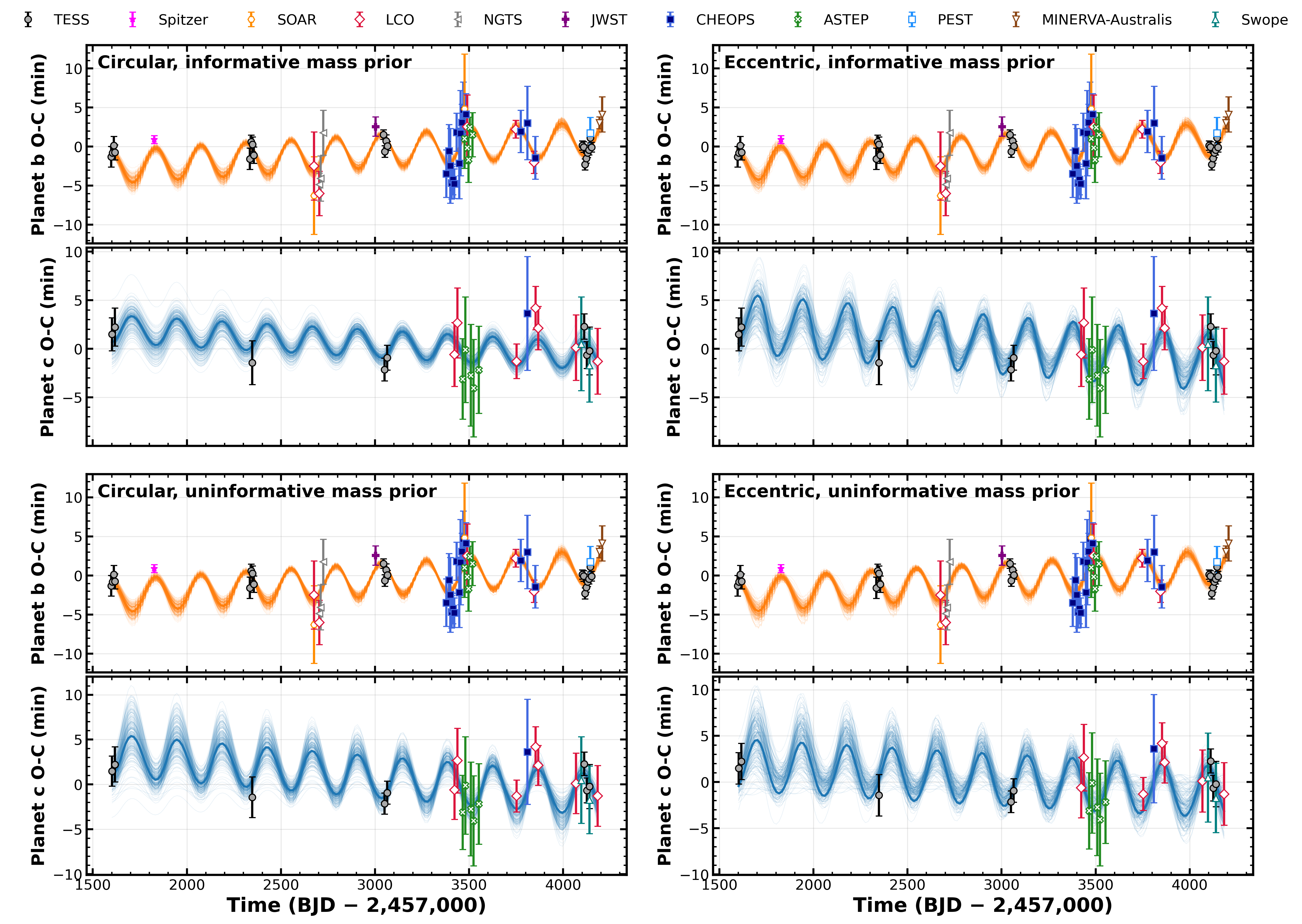}  \caption{O-C data (points) and models (orange for planet b and blue for planet c) for all four fits (circular on the left and eccentric on the right, Gaussian prior $M_b$ on top and uniform prior $M_b$ on bottom). The best-fit is a solid line, while the translucent lines are 100 random draws from the MCMC posterior. Points have colors and markers separated by instrument, with ground-based data hollow points and space-based data filled points. 
} \label{fig:OC}
\end{figure*}

\begin{figure}[htp]
    \includegraphics[width=0.48\textwidth]{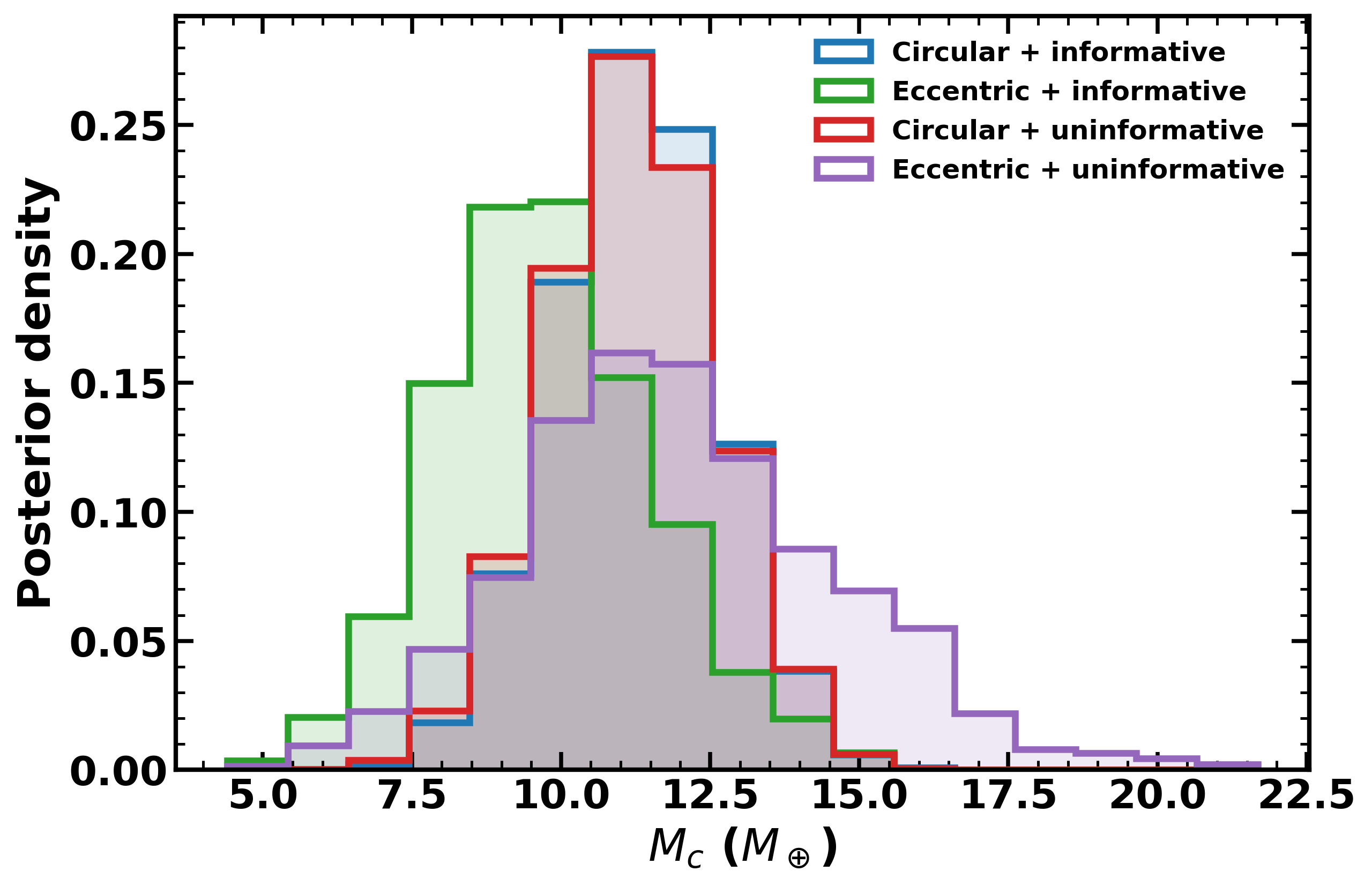}  \caption{Posterior distributions of the mass of planet c ($M_c$) from each of the four fits, colored by the fit assumptions. All fits yield a consistent $M_c$ of $\simeq11M_\oplus$ although the uncertainties are sensitive to assumptions about $M_b$ and the planetary eccentricities (see Figure\,\ref{fig:corner}).
} \label{fig:M_c}
\end{figure}

\begin{figure}[htp]
    \includegraphics[width=0.48\textwidth]{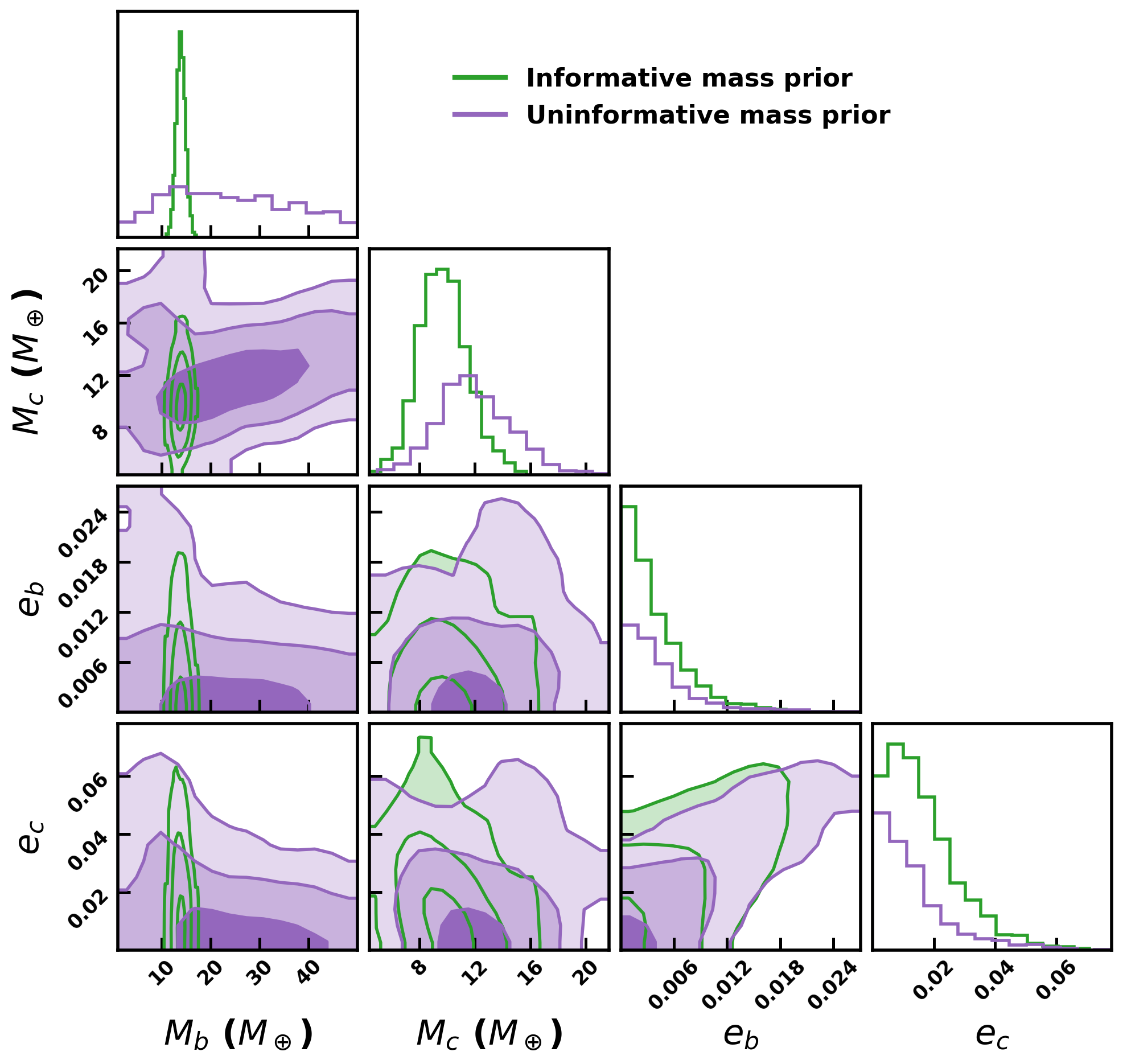}  \caption{Corner plot of the planet masses and orbital eccentricities for the fits where eccentricity was allowed to float, with an informative $M_b$ prior (green) or an uninformative $M_b$ prior (purple). Each contour level represents the 1$\sigma$, 2$\sigma$, and 3$\sigma$ values of the posterior output. The fit favors small eccentricities for both planets, although the two eccentricities are tightly correlated.
} \label{fig:corner}
\end{figure}

We report the median and 1$\sigma$ (16th and 84th percentile) values of the posteriors in Table\,\ref{tab:ttv_posteriors}. Figure\,\ref{fig:OC} shows the data and models for all four fits. We also show the mass posteriors for $M_c$ in Figure\,\ref{fig:M_c} and a corner plot of the major parameters in the two eccentric fits in Figure\,\ref{fig:corner}.

Table\,\ref{tab:ttv_posteriors} also includes statistics on the quality of the fits ($\chi^2_\nu$, $\Delta \chi^2$, and $\Delta {\rm BIC}$). The reduced $\chi^2$ values ($\chi^2_{\nu}$) for the four fits are $<1.0$; none of these are significant given the degrees of freedom. However, there is some evidence that the timing errors for the ground-based data are somewhat overestimated, mainly that the data can resolve some structure above the errorbars. For example, the flattening in c and the arc upward for b around TBJD of 3500 is visible in the ASTEP data despite relatively large timing errors. This likely arises from the GP over-fitting the data slightly. 

The data strongly rules out a scenario with no TTV. To quantify this, we compared the simplest TTV model (without $e$ and $\omega$) to a simple linear ephemeris for each planet. A linear ephemeris fit gives $\chi^2 \simeq135.7 $ for 74 degrees of freedom. The dynamical (\texttt{TTVfast}) fit improves this to $\chi^2 = 67.9$ for 72 degrees of freedom, an improvement of $\Delta\chi^2 = 67.8$ for 2 additional parameters ($M_b$ and $M_c$). Under the null hypothesis of no TTVs, the probability of obtaining such an improvement by chance is $2 \times 10^{-15}$, equivalent to a $\simeq8\sigma$ detection. The Bayesian Information Criterion (BIC) provides similarly strong evidence for the dynamical model ($\Delta\mathrm{BIC} = 59.1$). Therefore, despite the modest amplitude signal (peak-to-trough $\sim 5$\,minutes), the detection is robust. 

Assuming a circular orbit and a Gaussian $M_b$ prior (Fit\,1) yields a mass of $11.3\pm1.4M_\oplus$ for planet c, which is mostly independent of our assumptions about $M_b$. $M_b$ is more poorly constrained by the TTV data alone when allowed to float with uniform bounds (Fit \,3). The data favor a mass ($25.0^{+7.6}_{-7.8}\,M_\oplus$) intermediate to, and consistent with, both the lower mass from JWST transmission spectrum ($13.8 \pm 1.0\,M_\oplus$) and the higher mass ($27.7^{+5.9}_{-5.5}\,M_\oplus$) from VLT/CRIRES+ reported in \citet{Lavail2026}. 

When we remove the eccentricity restriction and adopt the transmission mass as a Gaussian prior (Fit\,2), the fit yields $M_c=9.7^{+1.9}_{-1.6}M_\oplus$ and low eccentricities for both planets (95$\%$ upper limits of $e_{b}<0.011$ and $e_{c}<0.039$). This is consistent with the circular-orbit fits. Indeed, models with $e\simeq0$ provide some of the highest likelihoods. $M_b$ is dominated by the prior, further indicating that the TTV is providing weaker constraints on the $M_b$ overall.

The final fit (Fit\,4), using a uniform $M_b$ prior and the KDE prior on $e$ and $\omega$, favors masses with larger uncertainties, but still in $\sim1\,\sigma$ agreement with the remaining 3 fits. $M_b$ again favors an intermediate mass to \citet{Thao2024_featherweight} and \citet{Lavail2026}, though the large uncertainties are unable to rule out either solution. The fit yields low eccentricities for both planets (95$\%$ upper limits of $e_{b}<0.014$ and $e_{c}<0.040$) and a reasonably constrained consistent $M_c$ ($11.8^{+3.0}_{-2.3}$\,M$_\oplus$).

\begin{figure}[htp]
    \centering
    \hspace*{-1cm}
    \includegraphics[width=0.54\textwidth]{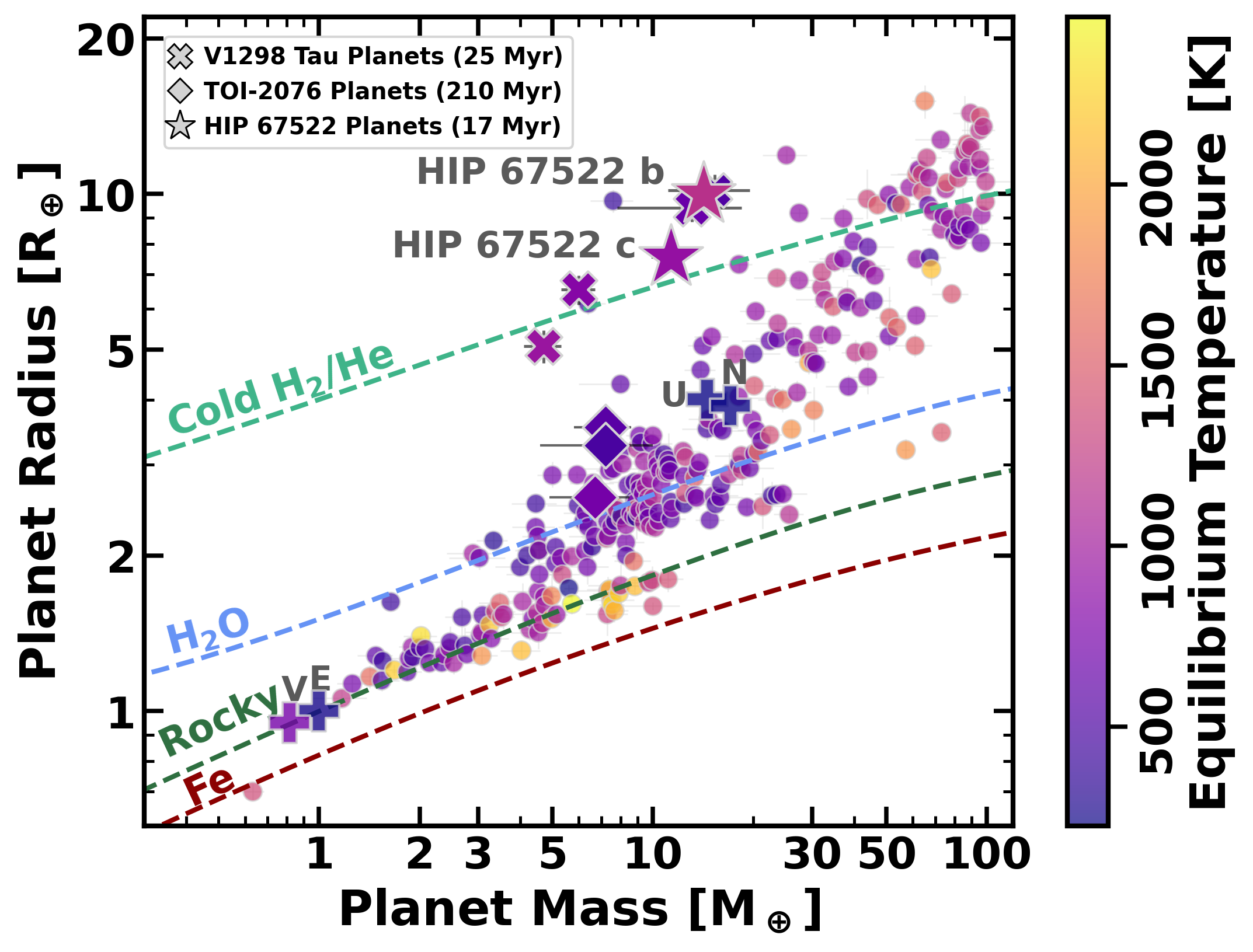}  \caption{Mass-radius distribution of exoplanets (circles) compared to young planets (V1298\,Tau bcde; Xs, HIP\,67522\,bc; stars, and TOI-2076 bcd; diamonds). Points are colored by their estimated equilibrium temperature. Masses and radii for the parent distribution were taken from the NASA exoplanet archive \citep[restricting to 20$\%$ precision cuts on both mass and radius;][]{NASAexplanetarchive}, V1298 Tau bcde are from \citet{Livingston2026}, and TOI-2076 bcd are from \citet{Wang2026_2076ttv}. The youngest planets (HIP\,67522 and V1298 Tau) have masses that show a clear offset from the parent distribution while the adolescent planets \citep[210\,Myr TOI-2076;][]{Barber2025_toi2076e} land closer to the mature sequence. The youngest planets clearly stand out as their own population and suggest they will likely evolve almost vertically into super-Earths/sub-Neptunes over the next $\simeq$Gyr. 
} \label{fig:mr_plot}
\end{figure}

Eccentricities need not be exactly zero even if they are low, and the b mass constraints from the transmission data are superior to those from our TTV analysis. As a result, we adopt the eccentric, Gaussian $M_b$ fit (Fit\,2) as the most reliable. Using these masses, we show how the masses and radii of HIP\,67522\,bc compare to the similar young planetary system V1298 Tau bcde and the population of (mostly) older planets in Figure\,\ref{fig:mr_plot}.

\begin{deluxetable*}{lcccc}
\tablecaption{Posterior parameter estimates for HIP\,67522\,b and c
              from four TTV fit configurations. \label{tab:ttv_posteriors}}
\tablewidth{0pt}
\tabletypesize{\footnotesize}
\tablehead{
  \colhead{} &
  \colhead{Fit\,1: Circular,}    & \colhead{Fit\,2: Eccentric,\tablenotemark{a}} & \colhead{Fit\,3: Circular,}  & \colhead{Fit\,4: Eccentric,\tablenotemark{a}} \\
  \colhead{Parameter} &
  \colhead{Gaussian $M_b$} & \colhead{Gaussian $M_b$} & 
  \colhead{Uniform $M_b$}  & \colhead{Uniform $M_b$}
}
\startdata
\sidehead{\bf Planet b}
$P_{\rm b}\tablenotemark{b}$ (d) & $6.959192 \pm 0.000033$ & $6.959283^{+0.000062}_{-0.000066}$ & $6.959194 \pm 0.000033$ & $6.959189^{+0.000088}_{-0.000101}$ \\
$T_{0,{\rm b}}$ (BJD$-$2,457,000) & $1597.05783 \pm 0.00083$ & $1597.0594 \pm 0.0012$ & $1597.05787 \pm 0.00087$ & $1597.0579^{+0.0015}_{-0.0017}$ \\
$M_{\rm b}$ ($M_\oplus$)          & $14.19 \pm 0.96$ & $13.97 \pm 0.95$ & $25.0^{+7.6}_{-7.8}$ & $23.1^{+15.4}_{-12.4}$ \\
$\sqrt{e_{\rm b}}\cos{\omega_{\rm b}}$ & $\equiv 0$ & $-0.014^{+0.052}_{-0.044}$ & $\equiv 0$ & $-0.0002^{+0.060}_{-0.044}$ \\
$\sqrt{e_{\rm b}}\sin{\omega_{\rm b}}$ & $\equiv 0$ & $-0.005^{+0.036}_{-0.041}$ & $\equiv 0$ & $0.020^{+0.039}_{-0.037}$ \\
$e_{\rm b}$   & $\equiv 0$ & $0.0029^{+0.0044}_{-0.0021}$ & $\equiv 0$ & $0.0033^{+0.0043}_{-0.0023}$ \\
\sidehead{\bf Planet c}
$P_{\rm c}$\tablenotemark{b} (d) & $14.33677 \pm 0.00012$ & $14.33649^{+0.00021}_{-0.00024}$ & $14.33817^{+0.00097}_{-0.00100}$ & $14.3378^{+0.0022}_{-0.0015}$ \\
$T_{0,{\rm c}}$ (BJD$-$2,457,000) & $1602.5161 \pm 0.0012$ & $1602.5144 \pm 0.0017$ & $1602.5275^{+0.0080}_{-0.0082}$ & $1602.524^{+0.016}_{-0.012}$ \\
$M_{\rm c}$ ($M_\oplus$)          & $11.3 \pm 1.4$ & $9.7^{+1.9}_{-1.6}$ & $11.2 \pm 1.4$ & $11.8^{+3.0}_{-2.3}$ \\
$\sqrt{e_{\rm c}}\cos{\omega_{\rm c}}$ & $\equiv 0$ & $0.063^{+0.043}_{-0.071}$ & $\equiv 0$ & $0.055^{+0.062}_{-0.078}$ \\
$\sqrt{e_{\rm c}}\sin{\omega_{\rm c}}$ & $\equiv 0$ & $-0.077^{+0.068}_{-0.062}$ & $\equiv 0$ & $-0.023^{+0.070}_{-0.075}$ \\
$e_{\rm c}$                       & $\equiv 0$ & $0.0135^{+0.0137}_{-0.0087}$  & $\equiv 0$ & $0.0103^{+0.0126}_{-0.0077}$ \\
\sidehead{\bf Fit statistics}
DoF                               & 72 & 68 & 72 & 68 \\
$\chi^2/$DoF                & 0.943 & 0.933 & 0.917 & 0.947 \\
$\Delta\chi^2$ vs.\ linear        & 67.8 & 72.3 & 69.7 & 71.3 \\
$\Delta\mathrm{BIC}$ vs.\ linear  & 59.1 & 46.1 & 60.9 & 45.1 \\
\enddata
\tablenotetext{a}{Sampled in $(\sqrt{e}\cos\omega,\,\sqrt{e}\sin\omega)$
  with a KDE prior derived from our \texttt{Juliet} fit (Section\,\ref{sec:juliet}).}
\tablenotetext{b}{\ttvcode{} uses a $t_{\rm ref}$ value which can differ from the traditional values observers use by $\simeq$minutes/orbit (e.g., due to precession). For future observation planning, we suggest using the values in Table\,\ref{tab:ephemeris}.}
\tablecomments{Values are posterior median with errors from the
  16th and 84th percentiles. The Gaussian mass prior on $M_{\rm b}$
  is $13.8 \pm 1.0\,M_\oplus$ from \citet{Thao2024_featherweight}; uniform mass priors
  are bounded at $[1,\,50]\,M_\oplus$. Eccentricity is fixed
  ($e \equiv 0$, $\omega$ undefined) in the circular-orbit fits.}
\end{deluxetable*}

\section{Summary and Discussion}\label{sec:summary}

HIP\,67522\,bc orbit near the 2:1 resonance and therefore are expected to exhibit TTVs due to mutual interactions between them. To characterize these TTVs, and measure the masses and eccentricities of the two planets, we collected \totaltransits\ transits of the HIP\,67522 system (\btransit{} of b and \ctransit{} of c across \bepochs\ epochs for b and \cepochs\ for c). Transits were drawn from eleven different facilities, including 4 space-based facilities (\tess, \jwst, \spitzer, and \cheops). We fit the \tess\ transits with a global model (allowing for transit timing variations) which we used as priors to sequentially fit the remaining transits. The timing results from these fits are generally quite precise, providing sub-minute timings in the best cases, and 4-6\,min timings in the worst cases.

Using \texttt{TTVFast}, we detect a significant TTV signal in both b and c. The peak-to-trough amplitude is small, about 5\,minutes for both b and c, but a TTV fit is highly favored over a linear ephemeris for both planets. The low amplitude is reflective of the low masses ($M_b=13.97\pm0.95\,M_\oplus$, $M_c=9.7^{+1.9}_{-1.6}\,M_\oplus$) and eccentricities (95$\%$ upper limits of $e_{b}<0.011$ and $e_{c}<0.039$) found, and the orbital periods just outside the true 2:1 resonance.

\subsection{Comparison to V1298 Tau and TOI 2076}

Multiple theories of planet formation attempt to reconstruct the \kepler\ small planet population \citep[e.g.,][]{Lee2014,Luque2022}. One such hypothesis, gas-dwarfs, suggests super-Earths and sub-Neptunes form from rocky cores with large, puffy atmospheres. These primordial envelopes are stripped, causing the observed planet radius to contract and settle into its mature size. Thus, young small planet progenitors are expected to be inflated and low-density \citep[e.g.,][]{Rogers2021}.

\plnameb\ and c have low densities offset from the mature planet population (see Figure\,\ref{fig:mr_plot}). The TTV analysis by \citet{Livingston2026} showed that the four $\sim$25\,Myr V1298 Tau planets \citep{David2019_v1298tau} are similarly low-mass despite their large radii. Together, these six infant planets form a preliminary mass-radius sequence inconsistent with the mature population but consistent with expectations for gas dwarf small-planet progenitors \citep[e.g.,][]{Rogers2025}. While some mature super-puffs occupy this regime \citep[e.g., TOI-791;][]{Dransfield2026_791}, they are rare, and all inflated young planets thus far with TTV-based masses (6 across HIP\,67522 and V1298 Tau) are low mass.

The next youngest system with a TTV-mass determination is the $\sim$210\,Myr TOI-2076 \citep{Wang2026_2076ttv}. The system hosts three sub-Neptunes in MMR \citep{Hedges2021_2706_1807, Osborn2022} and one inner super-Earth disjoint from the outer system \citep{Barber2025_toi2076e}. The outer three sub-Neptunes show significant TTVs. While the innermost planet is consistent with a stripped planet core, the outer three planets sit on the low-mass-edge of the mature planet population. By $200-500$\,Myr, gas dwarf models predict planets have largely finished contracting and are expected to shrink by $\lesssim1$R$_\oplus$ before maturity. This is consistent with what is seen here -- the sub-Neptune planets sit on the outskirts of the mature population and only minor changes to their radii would move them to the bulk distribution of the mature mass-radius sequence. 

Further work determining masses for the youngest planets, such as with TTVs, will be essential for mapping out the young planet mass-radius sequence and determining if the preliminary sequence seen here is statistically significant.

\subsection{Eccentricity and migration}

 Because of the age, the system has not had time for orbital evolution, so the measured eccentricities should be near-identical to the time of disk dispersal. The low values are consistent with expectations for Type I disk migration and resonant capture. In Type I disk migration, any induced eccentricity is quickly damped before the disk is dispersed \citep[e.g.,][]{Papaloizou2000_diskmige, Tanaka2004_diskmige}. For resonant capture, the outer planet migrates inward at a faster rate than the inner planet, where it is then `captured' at a first-order resonance. Interactions with the disk maintain low eccentricities \citep[$<0.1$; e.g.,][]{Lee2002_rescapture, Goldreich2014_rescapture}. While just one system, the eccentricity results here add to the small number of young planets with eccentricity measurements that can be used to test theories of migration.

\subsection{Comparison to Chakraborty et al. (2026)}

During the preparation of this manuscript, \citet{Chakraborty2026} published an analysis of the TTVs in the HIP\,67522 system using a subset of the transits included in this work. A check of the common transits in our analyses (from \tess\ and \cheops) suggests our methodologies produce consistent transit timings with similar uncertainties. Indeed, the \cheops\ transits in both analyses show a similar upward arc (at TBJD $\simeq$3500), and the minor differences in the \tess\ timing precisions likely arise from differences in light curve extraction (SPOC vs custom) and cadence (2-minute vs 20-second). 

\citet{Chakraborty2026} finds the TTVs in planet b to be consistent with a flat-line model (no TTVs exceeding a two-minute amplitude), while we find the TTV model is heavily preferred. This is in large part due to the sampling of transits presented in \citet{Chakraborty2026}; the analysis is largely reliant on the transits from \tess\ Sectors\,11, 38, and 64, all of which occur in a similar phase of the TTV model (see Figure\,\ref{fig:OC}) because the gap between successive windows is close to an integer multiple of the super-period of the system. Changes in the observing strategy of \tess{} mean that later epochs (including \tess\ Sectors\,101 and 102) sample a wider range of phases and the TTV signal becomes more apparent. The other difference is the wealth of ground-based data presented here, which covers more of the TTV period than is possible with the short \tess{} sectors. 

Consistent with this, the final mass limit reported in \citet{Chakraborty2026} for planet c ($<22M_\oplus$) is in agreement with ours. With additional transits, we were able to get a full TTV detection, allowing us to better map out the mass of planet c instead of just an upper limit.  

\subsection{Impacts of Spot Crossings and Depth Variations}

Some transits show visible spot crossings (e.g. Figure\,\ref{fig:lcogt}a). Visual inspection confirmed that the GP handled these well, leaving no significant structure in the residuals. These were also generally far from ingress/egress, and less likely to impact the timings. As a further test, we tried fitting these with a simple Gaussian and found no advantage over the global GP model (times differed by at most $<0.5\sigma$). This matches prior studies showing that spot-induced timing variations are negligible when fitting with a GP \citep{LopezMurillo2026}, and instead come up when using a simple polynomial model and ignoring the spot crossings \citep{Mazeh2015a}.

While we do not explicitly model transit depth variations, we find that our transit depths are generally consistent across the entire dataset. This is largely due to our decision to place priors on $R_{\rm p}/R_*$ from the \tess{} data. Indeed, we can see in some transits (Figure\,\ref{fig:cheops_full}) that the GP is `dipping' into the transit, perhaps to maintain the $R_{\rm p}/R_*$ in the prior. While \citet{Chakraborty2026} does find significant transit depth variations (peak-to-peak amplitude $>$30\%), our transit-fitting methodology is not ideal for testing questions about transit depth variations.

\subsection{Potential for constraining $M_b$}

The unconstrained $M_b$ fits yield a mass for planet b consistent with masses derived from the JWST transmission spectrum ($\sim$14\,M$_\oplus$) and the CRIRES+ transmission spectrum ($\sim$28\,M$_\oplus$). The least restrictive fit here, allowing for eccentric orbits and a uniform prior on $M_b$ (Fit\,4), finds $M_b$ to be intermediate between the two ($\sim$23\,M$_\oplus$) and both solutions are consistent at the 1$\sigma$ level. The result is most limited by the smaller sample of planet c transits. Future work with additional transits of planet c will be able to more tightly constrain the mass of planet b through TTVs alone, as the case for planet c. However, the results here still lend credence to masses derived from transmission spectroscopy. This fits with results on V1298\,Tau b, where the transmission spectrum favors a mass consistent with the TTV signal \citep{Barat2025_jwst, Livingston2026}. More targets and more precise mass constraints are needed to further validate masses from transmission spectra, especially for young planets where RVs are more challenging.

\begin{acknowledgments}

HJ acknowledges funding from a Summer Undergraduate Research Fellowship from the Office for Undergraduate Research at the University of North Carolina at Chapel Hill. MGB was supported by the NSF Graduate Research Fellowship (DGE-2040435) and the North Carolina Space Grant Graduate Research Fellowship Program. AWM was supported by grants from the TESS Guest investigator program (80NSSC25K0113), the NSF CAREER program (AST-2143763), and NASA's Exoplanet Research Program (XRP 80NSSC25K7148). PCT was supported by NSF Graduate Research Fellowship (DGE-1650116), the North Carolina Space Grant Graduate Research Fellowship Program, and a grant from the TESS Guest investigator program (80NSSC24K1142). AILM was supported by a grant from the TESS guest investigator program (80NSSC25K7903) and the North Carolina Space Grant Graduate Research Fellowship Program. This work is partly supported by JSPS KAKENHI Grant Numbers JP24H00017, JP24K17083, JP25K24620, JP26H01402, JP26K00755, JP24K00689, and JP25K17450, and JSPS Grant-in-Aid for JSPS Fellows Grant Number JP24KJ0241. 

This work makes use of observations from the Las Cumbres Observatory global telescope network.

This paper is based on observations made with the MuSCAT3/4 instruments, developed by the Astrobiology Center (ABC) in Japan, the University of Tokyo, and Las Cumbres Observatory (LCOGT). MuSCAT3 was developed with financial support by JSPS KAKENHI (JP18H05439) and JST PRESTO (JPMJPR1775), and is located at the Faulkes Telescope North on Maui, HI (USA), operated by LCOGT. MuSCAT4 was developed with financial support provided by the Heising-Simons Foundation (grant 2022-3611), JST grant number JPMJCR1761, and the ABC in Japan, and is located at the Faulkes Telescope South at Siding Spring Observatory (Australia), operated by LCOGT.

This work makes use of observations from the ASTEP telescope. ASTEP benefited from the support of the French and Italian polar agencies IPEV and PNRA in the framework of the Concordia station program, from OCA, INSU, ANR (EXTRASTEP) and ESA through the Science Faculty of the European Space Research and Technology Centre (ESTEC). This research is also is supported by the European Union's Horizon 2020 research and innovation programme (grant's agreement n$^{\circ}$ 803193/BEBOP), and from the Science and Technology Facilities Council (STFC; grant n$^\circ$ ST/S00193X/1, ST/W002582/1, and ST/Y001710/1).

These observations are associated with program JWST-GO-02498. Support for program JWST-GO-02498 was provided by NASA through a grant from the Space Telescope Science Institute, which is operated by the Association of Universities for Research in Astronomy, Inc., under NASA contract NAS 5-03127. This work is based (in part) on observations made with the NASA/ESA/CSA James Webb Space Telescope. The data were obtained from the Mikulski Archive for Space Telescopes at the Space Telescope Science Institute, which is operated by the Association of Universities for Research in Astronomy, Inc., under NASA contract NAS 5-03127 for JWST. These observations are associated with program \#2498.

This paper includes data collected with the TESS mission, obtained from the MAST data archive at the Space Telescope Science Institute (STScI). Funding for the TESS mission is provided by the NASA Explorer Program. STScI is operated by the Association of Universities for Research in Astronomy, Inc., under NASA contract NAS 5–26555.

This work is based [in part] on archival data obtained with the Spitzer Space Telescope, which was operated by the Jet Propulsion Laboratory, California Institute of Technology under a contract with NASA. Support for this work was provided by an award issued by JPL/Caltech.

Based on observations obtained at the Southern Astrophysical Research (SOAR) telescope, which is a joint project of the Minist\'{e}rio da Ci\^{e}ncia, Tecnologia e Inova\c{c}\~{o}es (MCTI/LNA) do Brasil, the US National Science Foundation’s NOIRLab, the University of North Carolina at Chapel Hill (UNC), and Michigan State University (MSU)

\end{acknowledgments}

\facilities{ASTEP, LCOGT, SOAR, \jwst, \tess, \spitzer, \cheops,  MINERVA-Australis, Swope, NGTS, PEST}

\software{
\texttt{LDTK} , \texttt{emcee}, \texttt{corner.py} \citep{foreman2016corner}, \texttt{celerite}, \texttt{matplotlib} \citep{hunter2007matplotlib}, \texttt{batman}, \texttt{Astropy} \citep{Astropy2013, AstropyCollaboration2018, Astropy2022}, \texttt{numpy} \citep{Harris2020}, \texttt{juliet} \citep{espinoza2019juliet}, \texttt{TTVFast}  \citep{Speagle2020_dynesty,Dynesty_2023} 
}

\begin{deluxetable*}{lcccccr} 
\tabletypesize{\footnotesize} 
\tablecaption{Transit observations and $T_c$ measurements analyzed in this work}\label{tab:obslog}

\tablewidth{0pt}
\tablehead{
\colhead{Epoch} &
\colhead{Start Date (UT)} &
\colhead{Telescope} & 
\colhead{Filter} &
\colhead{Exp time (s)} &
\colhead{Coverage} & 
\colhead{$T_{C}$ (BJD)}}
\startdata
     \multicolumn{7}{c}{\textbf{Planet b}}\\ 
     \hline
     0 & 2019-04-23 & \tess\, Sector 11 & \tess\ & 120 & Full & $2458597.06376 \pm 0.00091$ \\
     1 & 2019-04-30 & \tess\, Sector 11 & \tess\ & 120 & Full & $2458604.02358 \pm 0.00054$ \\
     2 & 2019-05-07 & \tess\, Sector 11 & \tess\ & 120 & Full & $2458610.98371 \pm 0.00078$ \\
     3 & 2019-05-14 & \tess\, Sector 11 & \tess\ & 120 & Full & $2458617.94256 \pm 0.00067$ \\
     33 & 2019-12-09 & \spitzer\, &  IRAC 4.5 & 1.9 & Full & $2458826.72779 \pm 0.00035$ \\
     106 & 2021-04-30 & \tess\, Sector 38 & \tess\, & 120  &  Full & $2459334.76738 \pm 0.00095$  \\
     107 & 2021-05-07 & \tess\, Sector 38 & \tess\, & 120  &  Full & $2459341.72836 \pm 0.00060$ \\
     108 & 2021-05-14 & \tess\, Sector 38 & \tess\, & 120  &  Full & $2459348.68760 \pm 0.00066$ \\
     109 & 2021-05-21 & \tess\, Sector 38 & \tess\, & 120  &  Full & $2459355.64610 \pm 0.00073$ \\
     155 & 2022-04-06 & SOAR & SDSS $g'$ & 8.25 & Full & $2459675.77814 \pm 0.00344$ \\
     \multicolumn{7}{c}{...}\\
    \hline
    \multicolumn{7}{c}{\textbf{Planet c}}\\ 
    \hline 
    0 & 2019-04-29 & \tess\, Sector 11  & \tess\, & 120  & Full & $2458602.50211 \pm 0.00117$ \\
    1 & 2019-05-13 & \tess\, Sector 11  & \tess\, & 120  & Full & $2458616.83756 \pm 0.00135$ \\
    52 & 2021-05-13 & \tess\, Sector 38 & \tess\, & 120  &  Full & $2459347.91620 \pm 0.00156$ \\
    101 & 2023-04-15 & \tess\, Sector 64 & \tess\, & 20  &  Full & $2460050.32703 \pm 0.00081$ \\
    102 & 2023-04-30 & \tess\, Sector 64 & \tess\, & 20  &  Full & $2460064.66281 \pm 0.00088$ \\
    127 & 2024-04-22 & MuSCAT4 & $g_{p}, r_{p}, i_{p}, z_{s}$ & 12, 7, 8, 5 & Partial & $2460423.03615 \pm 0.00228$ \\
    128 & 2024-05-06 & LCO & $r_{p}$ & 10 & Partial & $2460437.37334 \pm 0.00250$ \\
    130 & 2024-06-04 & ASTEP & ASTEP & 2 & Full & $2460466.03919 \pm 0.00287$ \\
    131 & 2024-06-18 & ASTEP & ASTEP & 2 & Full & $2460480.37620 \pm 0.00377$ \\
    \multicolumn{7}{c}{...}\\
\enddata
\tablecomments{A full version of this table is available from the journals of the AAS as a Machine-Readable Table. }
\end{deluxetable*}
\clearpage

\bibliographystyle{apj}
\bibliography{Mann_Toddlers, planetSearch}
\end{document}